\documentclass[english,11pt,a4paper]{article}
\usepackage[T1]{fontenc}
\usepackage[latin9]{inputenc}
\usepackage{amsmath}
\usepackage{amsthm}
\usepackage{amssymb}

\usepackage{tikz}
\usetikzlibrary{decorations.markings}
\usetikzlibrary{arrows}
\usetikzlibrary{tikzmark,calc}
\usepackage{relsize}

\makeatletter
\numberwithin{equation}{section}

\usepackage{jheppub}
\usepackage{slashed}
\usepackage{fontawesome5}
\allowdisplaybreaks

\makeatother

\usepackage{babel}
\begin{document}
\title{A cosmological sandwiched window for lepton-number breaking scale}
\abstract{A singlet majoron can arise from the seesaw framework as a pseudo-Goldstone boson when the heavy Majorana neutrinos acquire masses via the spontaneous breaking of global ${\rm U}(1)_L$ symmetry. 
The resulting cosmological impacts are usually derived from the effective majoron-neutrino interaction, and the majoron abundance is accumulated through the freeze-in neutrino coalescence. However, a primordial majoron abundance can be predicted in a minimal setup and lead to distinctive cosmological effects. In this work, we consider such a primordial majoron abundance from relativistic freeze-out and calculate the modification to the effective neutrino number $N_{\rm eff}$.  We demonstrate that the measurements of $N_{\rm eff}$ will constrain the parameter space from a primordial majoron abundance in an opposite direction to that from neutrino coalescence. When the contributions from both the primordial abundance and the freeze-in production coexist, the ${\rm U}(1)_L$-breaking scale (seesaw scale) $f$ will be pushed into a ``sandwiched window''. 
Remarkably, for majoron masses below
1 MeV and above the eV scale, the future CMB-S4 experiment will completely close such a low-scale seesaw window for $f\in [1,10^5]~{\rm GeV}$. We highlight that any new light particle with a primordial abundance that couples to Standard Model particles may lead to a similar sandwiched window, and such a general phenomenon deserves careful investigation.
}

\author[a]{Shao-Ping Li} 
\author[a,b]{and Bingrong Yu} 
\affiliation[a]{Institute of High Energy Physics, Chinese Academy of Sciences, Beijing 100049, China} 
\affiliation[b]{Department of Physics, LEPP, Cornell University, Ithaca, NY 14853, USA}
\preprint{\today}  
\emailAdd{spli@ihep.ac.cn} 
\emailAdd{bingrong.yu@cornell.edu}  

\definecolor{color_git}{rgb}{0.098, 0.160, 0.345}
\newcommand{\gitlink}{\href{https://github.com/Shao-Ping-Li}{\textsc{g}it\textsc{h}ub {\large\color{color_git}\faGithub}}} 
\newcommand{\nudec}{\href{https://github.com/MiguelEA/nudec_BSM}{$\textsc{NUDEC}_{-}$\textsc{BSM} {\large\color{color_git}\faGithub}}} 

\newcommand{\SP}[1]{{\color{blue} #1}}

\preprint{\today}
\maketitle

\section{Introduction}
Over the past few decades, cosmological observations have reached unprecedented precision, which offers a promising avenue for probing new physics beyond the Standard Model (SM). For instance, nowadays  the measurements of the effective number of neutrino species $N_{\rm eff}$ have reached the accuracy of ${
\cal O} (0.1)$ through the observations of big bang nucleosynthesis (BBN) and cosmic microwave background (CMB)~\cite{Planck:2018vyg,Yeh:2022heq}. In general, any new light particles coupling to the SM particles might contribute to the relativistic degrees of freedom in the early universe and lead to significant deviations from the SM prediction of $N_{\rm eff}$, and therefore will receive strict constraints from cosmology.

In this work, we revisit the majoron-neutrino interactions under the cosmological constraints of $N_{\rm eff}$ and study their possible implications for the low-scale seesaw scenario. A singlet majoron $J$ can naturally arise as a pseudo-Goldstone boson~\cite{Chikashige:1980ui} in the framework of the type-I seesaw model~\cite{Minkowski:1977sc, Yanagida:1979as, Gell-Mann:1979vob, Glashow:1979nm, Mohapatra:1979ia}, where the heavy Majorana neutrinos acquire masses through the spontaneous breaking of the global ${\rm U}(1)_L$ lepton-number symmetry. The lepton-number breaking scale $f$ characterizes the mass scale of heavy neutrinos, i.e., the scale of new physics responsible for the origin of neutrino masses. After the electroweak gauge symmetry breaking, the majoron will couple with the SM neutrinos $\nu_i$ through the mixing between active and sterile neutrinos, where the couplings are suppressed by $m_i/f$ (with $m_i$ the mass of $\nu_i$). Therefore, the majoron is expected to be long-lived due to the large hierarchy between $m_i$ and $f$, which makes the majoron
an interesting candidate of dark matter (DM) \cite{Berezinsky:1993fm,Lattanzi:2007ux,Frigerio:2011in,Bazzocchi:2008fh,Lattanzi:2013uza,Queiroz:2014yna,Garcia-Cely:2017oco,Akita:2023qiz,Xu:2023xva}.\footnote{If the ${\rm U}(1)_L$ global symmetry is only broken spontaneously, then the majoron will be strictly massless as a real Nambu-Goldstone boson. However, to serve as a DM candidate, the majoron should acquire a nonzero mass where the global symmetry is broken explicitly~\cite{Langacker:1986rj,Akhmedov:1992hi,Rothstein:1992rh,Gu:2010ys,Frigerio:2011in}.} For a relatively low breaking scale $f\lesssim 100~{\rm TeV}$, which corresponds to the low-scale seesaw scenario, the decay of the majoron into a pair of SM neutrinos could occur and contribute to $N_{\rm eff}$ at some crucial epochs of the Universe, leaving observable imprints in, e.g., the BBN and CMB. Given the strict constraints on $N_{\rm eff}$ from current and upcoming CMB experiments~\cite{SimonsObservatory:2018koc,SimonsObservatory:2019qwx,Planck:2018vyg,CMB-S4:2016ple,Abazajian:2019eic}, it is hopeful that we could probe the mechanism for neutrino mass generation from observables in the early Universe.

The constraints on the lepton-number breaking scale from $N_{\rm eff}$ have been noticed for a while~\cite{Chacko:2003dt,Boehm:2012gr,Escudero:2019gvw,Sandner:2023ptm,Li:2023puz}, mostly in a model-independent setup.  In previous works, the primordial abundance of the majoron was usually neglected such that
the majoron abundance was initialized via the effective interaction between the active neutrinos and the majoron. In particular, for a light majoron below the MeV scale, the majoron abundance is accumulated through the freeze-in $2\nu\to J$ process where the majoron-neutrino coupling is sufficiently small. In this case,  the primordial majoron abundance is negligible before the SM neutrino decoupling, and the decay of the majoron into a pair of neutrinos $J\to 2\nu$ after the SM neutrino decoupling will inject energies into neutrinos, causing an excess of  $N_{\rm eff}$. Consequently, a smaller breaking scale $f$ (equivalent to a larger majoron-neutrino coupling) brings about a larger majoron abundance, thereby leading to a larger deviation of $N_{\rm eff}$. Therefore, the constraints from the $N_{\rm eff}$ measurements   will put    a lower bound on $f$  in the freeze-in situation.
 
However, a primordial majoron abundance generated beyond the effective majoron-neutrino interaction in general \emph{cannot} be neglected before the SM neutrino decoupling.  
In fact, it can be   predicted even in a minimal setup and more importantly, it may have a sizable effect to $N_{\rm eff}$.  In this work, we investigate such an effect that has usually been neglected in the literature. We focus on the situation where the majoron is assumed to be in thermal equilibrium with SM particles in the early Universe, such that  the primordial majoron abundance is  inherited from relativistic freeze-out.  As can be seen later, this situation is easy to realize, either via the interactions between the majoron and the heavy Majorana neutrinos or through the majoron-Higgs interactions in the scalar sector.  We will also consider the situation where the primordial majoron abundance is accumulated by some other mechanism beyond the relativistic freeze-out.  
  
The primordial majoron abundance will lead to several interesting phenomena. First of all, for a non-negligible primordial   abundance, if it exists before the SM neutrino decoupling but is only depleted into radiation near the recombination epoch,  $N_{\rm eff}$ will be drastically increased by orders of magnitude (see Sec.~\ref{sec:RfoNRdec}). Therefore, the precision measurements of $N_{\rm eff}$ at the CMB epoch will severely constrain the abundance,  and put strict bounds on the ultraviolet (UV)  physics that features the abundance. Furthermore, for nonrelativistic majoron decay, a larger breaking scale $f$ leads to  later decay, and hence a larger modification to $N_{\rm eff}$. So the constraints from $N_{\rm eff}$ will    put an upper bound on $f$, in contrast to the freeze-in situation. This is particularly interesting because when the freeze-in  and primordial abundances  coexist, the constraint from the $N_{\rm eff}$ measurements will  push $f$ into a sandwiched window. Remarkably, the next-generation CMB experiments could further narrow or   completely close such a sandwiched window~\cite{SimonsObservatory:2018koc,SimonsObservatory:2019qwx, CMB-S4:2016ple,Abazajian:2019eic}. 

The remaining part of this paper is organized as follows. In Sec.~\ref{sec:framework}, we start with a brief review of the singlet majoron model. Then we perform a general analysis of the majoron evolution in the early Universe and its cosmological impacts. In Sec.~\ref{sec:analytic_Neff}, we derive $N_{\rm eff}$ using an approximate analytical (but intuitive) method by assuming  instantaneous majoron decay. A     stricter calculation of $N_{\rm eff}$ beyond   instantaneous majoron decay is conducted in Sec.~\ref{sec:numerics}.
The constraints on $f$ from $N_{\rm eff}$ are given in Sec.~\ref{sec:constraints}, which provide a cosmological sandwiched window for the low-scale seesaw scenario. 
We summarize our main results in Sec.~\ref{sec:conclusion}. Finally, some technical details are included in the appendices.

\section{Framework}
\label{sec:framework}
\subsection{The singlet majoron model}
The singlet majoron model~\cite{Chikashige:1980ui} introduces a complex scalar $S$, which is a singlet under the SM gauge symmetries. The relevant Lagrangian in the Yukawa sector is given by
\begin{align}
{\cal L}= -\overline{\ell_{\rm L}}Y_\nu \widetilde{\Phi} N_{\rm R}-\frac{1}{2} \overline{N_{\rm R}^c}Y_N N_{\rm R} S + {\rm h.c.}\,,
\label{eq:Lag}
\end{align}
where $\ell_{\rm L}=\left(\nu_{\rm L}, l_{\rm L}\right)^{\rm T}$ and $\widetilde{\Phi}\equiv {\rm i}\sigma^2 \Phi^{*}$ are the left-handed lepton doublet and the Higgs doublet, respectively. $Y_\nu$ is the Dirac neutrino Yukawa coupling matrix, and  $Y_N$ is the Yukawa coupling matrix for the right-handed (RH) neutrino singlets $N_{\rm R}$, where $N_{\rm R}^c \equiv C \overline{N_{\rm R}}^{\rm T}$ has been defined with $C = {\rm i}\gamma^2 \gamma^0$ the charge-conjugation matrix. 

The Lagrangian in Eq.~(\ref{eq:Lag}) owns a global ${\rm U}(1)_{L}$  symmetry if we assign the lepton numbers of relevant particles to be: $L(\ell_{\rm L})=L(N_{\rm R})=+1$, $L(\Phi)=0$, and $L(S)=-2$. This global symmetry is spontaneously broken after $S$ acquires a non-zero vacuum expectation value (VEV) from the scalar potential (see Appendix~\ref{appen:majoron-Higgs} for the discussion about a general scalar potential obeying the lepton-number conservation)
\begin{align}\label{eq:VS}
V \supset\mu_S^2\left(S^\dagger S\right)+ \frac{\lambda_S}{2}\left(S^\dagger S\right)^2.
\end{align}
In this work, we adopt the linear realization of the broken symmetry. That is, the complex scalar is parametrized as\footnote{See Ref.~\cite{Frigerio:2011in} for the nonlinear realization of the broken ${\rm U}(1)_{L}$  symmetry, where $S$ is parametrized as $S=\left(\rho+f\right)\,e^{{\rm i}J/f}/\sqrt{2}$.}
\begin{eqnarray}
S = \frac{1}{\sqrt{2}}\left(f+\rho+{\rm i}J\right),
\end{eqnarray}
where $\rho$ and $J$ are two real degrees of freedom.  In addition, $f$ is the scalar VEV, corresponding to the lepton-number breaking scale, which provides a Majorana mass term $M_{\rm R}=Y_N f/\sqrt{2}$ for the RH neutrinos.  Note that  $f$ is also the seesaw scale provided ${\cal O}(Y_N) \simeq \mathcal{O}(1)$, i.e., $f \simeq {\cal O}(M_{\rm R})$, which is usually considered as the UV completion of the canonical seesaw model~\cite{Minkowski:1977sc, Yanagida:1979as, Gell-Mann:1979vob, Glashow:1979nm, Mohapatra:1979ia}.
 
From Eq.~(\ref{eq:VS}), it is easy to show that  
$\rho$ will acquire a mass proportional to $f$ after the lepton-number breaking while $J$ remains massless. Therefore, the pseudo-scalar $J$ (i.e., the majoron) is identified as the Goldstone boson of the spontaneous ${\rm U}(1)_{L}$ symmetry breaking. In practice, a non-zero majoron mass $m_J$ can be generated by adding terms that explicitly violate the ${\rm U}(1)_L$ symmetry,  at either tree or loop levels~\cite{Langacker:1986rj,Akhmedov:1992hi,Rothstein:1992rh,Gu:2010ys,Frigerio:2011in}. A simple case of mass generation is discussed  in Appendix~\ref{appen:majoron-Higgs}.  In the following analysis, we will simply treat the majoron mass as a free parameter. 

After the breaking of the electroweak gauge symmetry, the majoron can interact with active neutrinos through the flavor mixing between active and sterile neutrinos. The general interaction in the mass basis turns out to be~\cite{Pilaftsis:1993af}
\begin{eqnarray}
	\label{eq:LJnuij}
{\cal L}_J = -\frac{{\rm i}J}{2f}\sum_{i,j=1}^{6} \overline{n_i}\left[\gamma_5\left(m_i+m_j\right)\left(\frac{1}{2}\delta_{ij}-{\rm Re}\,{\cal C}_{ij}\right)+{\rm i}\left(m_i-m_j\right){\rm Im}\,{\cal C}_{ij}\right]n_j\;.
\end{eqnarray}
A detailed derivation is presented in Appendix~\ref{appen:majoron-neutrino}. 
In Eq.~(\ref{eq:LJnuij}), $n_i$ denotes the neutrino mass eigenstate with  mass  $m_i$, and the indices $i=1,2,3$ ($i=4,5,6$) correspond to the active (sterile) neutrino species. The mixing parameters ${\cal C}_{ij}\equiv \sum_{k=1}^{3}{\cal U}_{ki}^{*}{\cal U}_{kj}$  are defined by  the $6 \times 6$ unitary matrix  ${\cal U}$ which  diagonalizes the neutrino mass matrix via
\begin{eqnarray}
{\cal U}^\dagger 
\begin{pmatrix}
	0&M_{\rm D}\\
	M_{\rm D}^{\rm T}&M_{\rm R}
\end{pmatrix} {\cal U}^{*} = {\rm Diag}\left(m_1,m_2,...,m_6\right),
\end{eqnarray}
with $M_{\rm D}=Y_\nu v/\sqrt{2}$  the Dirac neutrino mass matrix and $v \simeq 246~{\rm GeV}$  the electroweak VEV. Note that ${\cal O}\left(M_{\rm D}\right) \ll  {\cal O}\left(M_{\rm R}\right)$ is required to generate the tiny masses for active neutrinos through the seesaw mechanism. 

To obtain the interaction between the majoron and the active neutrinos, one can simply take $i,j=1,2,3$ in Eq.~(\ref{eq:LJnuij}) and identify $n_i \equiv \nu_i$ (for $i=1,2,3$). Then it follows that ${\cal C}_{ij} \simeq \delta_{ij}$, where the non-diagonal elements are suppressed by ${\cal O}\left(M_{\rm D}^2/M_{\rm R}^2\right)$, leading to
\begin{eqnarray}
	\label{eq:Jnunu}
	{\cal L}_{J\nu\nu}\simeq \frac{{\rm i} J}{2f}\sum_{i=1}^3 m_i\overline{\nu_i}\gamma_5 \nu_i\;.
\end{eqnarray}
Therefore, the interaction between the majoron and the active neutrinos is approximately diagonal  and is suppressed by  feeble majoron-neutrino couplings $g_{\nu_i}\equiv m_i/f$.

For the mass region of $1~\text{eV}\lesssim m_J \lesssim 1~\text{MeV}$ which will be considered in this work, the majoron can decay to two active neutrinos, where the decay width is given by
\begin{align}
\Gamma_{J\to2\nu} =\frac{1}{16\pi f^2}\sum_{i=1}^{3}m_i^2\sqrt{m_J^2-4m_i^2} \simeq\frac{m_{J}}{16\pi f^{2}}\sum_{i=1}^{3}m_{i}^{2}\,.
\label{eq:Jto2nu}
\end{align}
In addition, the majoron can also decay to two photons at two-loop level with the width scaling as~\cite{Garcia-Cely:2017oco}
\begin{align}
\Gamma_{J\to 2\gamma}\sim \alpha^2 \text{Tr}\left(Y^{}_\nu Y^\dagger_\nu\right)^2m_J^3/f^2\,,
\end{align}
where $\alpha$ is the fine-structure constant. The di-photon decay mode is severely constrained by the CMB, X-, gamma- and cosmic-ray observations~\cite{Bazzocchi:2008fh,Lattanzi:2013uza,Garcia-Cely:2017oco}. In fact, for the low-scale seesaw scenario (i.e., $f\lesssim 100~{\rm TeV}$) considered in this paper, the Yukawa couplings $Y_\nu$ are suppressed by the tiny neutrino masses (recall that the seesaw relation gives $Y_\nu \sim \sqrt{m_i f}/v \lesssim 10^{-5}$), thereby making $J\to 2\nu$ the dominant decay mode of the majoron.

\subsection{Majoron cosmology in the low-scale seesaw scenario}  
 
\begin{figure}
	\centering
	
	\includegraphics[scale=0.7]{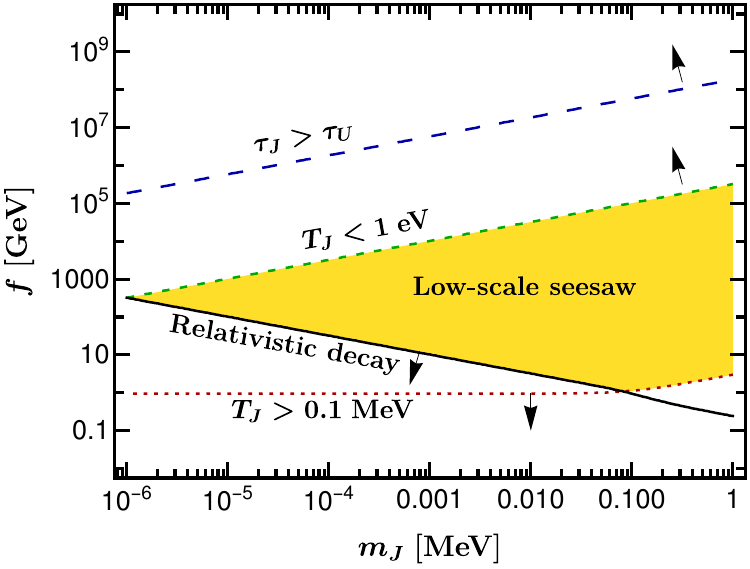} 
	
	\caption{\label{fig:DecayRegime} The decaying temperature of the majoron $T_J$ with different values of the majoron mass $m_J$ and the lepton-number breaking scale $f$. Some typical epochs  of the majoron decay are shown in the plot where:  the majoron is stable over the cosmological time scale $\tau_U\simeq 4.4\times 10^{17}~{\rm s}$ (above the blue long-dashed line),  the majoron decays after the epoch of the matter-radiation equality  $T_{\rm eq}\simeq 1~{\rm eV}$ (above the green short-dashed line), the majoron decays before the SM neutrino decoupling epoch $T_{\nu,{\rm dec}}\simeq 0.1~{\rm MeV}$ (below the red dotted line), and the majoron decays as a relativistic particle (below the black solid line). The yellow-shaded region is the parameter space which interests us in this work, i.e., the majoron decays nonrelativistically with the decaying temperature $T_{\rm eq} \lesssim T_J \lesssim T_{\nu, {\rm dec}}$. The lepton-number breaking scale in the yellow-shaded region resides in $1~{\rm GeV} \lesssim f \lesssim 100~{\rm TeV}$, which corresponds to the low-scale seesaw scenario.}
\end{figure}
 
Due to the large hierarchy between $m_i$ and $f$, the majoron is expected to be long-lived [cf. Eq.~(\ref{eq:Jto2nu})]. For a large enough $f$, the majoron can naturally serve as a DM candidate~\cite{Berezinsky:1993fm,Lattanzi:2007ux,Frigerio:2011in,Bazzocchi:2008fh,Lattanzi:2013uza,Queiroz:2014yna,Garcia-Cely:2017oco,Akita:2023qiz,Xu:2023xva}. For relatively small $f$, e.g., around the electroweak scale, the majoron becomes unstable within the cosmological time scale and will dominantly decays to active neutrinos. 
 
To see when the majoron decays, let us consider the temperature $T_J$ of nonrelativistic majoron decay, which can be defined as 
 \begin{align}\label{eq:TJ}
 	\tau_J^{-1}=\Gamma_J \simeq \Gamma_{J\to2\nu} \equiv 2H(T_J)\,.
 \end{align}
Here $H(T)\simeq 1.66\sqrt{g_{\rho}(T)}T^{2}/M_{{\rm Pl}}$ is the Hubble parameter at the radiation-dominated epoch, with $g_{\rho}(T)$ the relativistic degrees of freedom for energy density and $M_{{\rm Pl}}\simeq1.22\times10^{19}$~GeV the Planck mass. Moreover, the decay width of the majoron $\Gamma_J$ can be calculated by Eq.~(\ref{eq:Jto2nu}) as an approximation.
Note that for relativistic majoron decay, the lifetime $\tau_J$ in Eq.~\eqref{eq:TJ} is enhanced by an additional Lorentz factor $E_J/m_J>1$. Combining Eqs.~(\ref{eq:Jto2nu}) and (\ref{eq:TJ}), we obtain that the nonrelativistic decaying temperature $T_J$ scales as 
\begin{align}
	\label{eq:TJscale}
T_J &\simeq 0.077\, g_{\rho}^{-1/4} \frac{1}{f} \left(\sum_{i=1}^{3} m_i^2 m_J M_{\rm Pl}\right)^{1/2}\nonumber\\
&\simeq 0.997 \left(\frac{3.36}{g_{\rho}}\right)^{1/4} \left(\frac{\sqrt{\sum_{i=1}^{3} m_i^2}}{0.05~{\rm eV}}\right)\left(\frac{m_J}{0.1~{\rm MeV}}\right)^{1/2}\left(\frac{f}{100~{\rm GeV}}\right)^{-1}~{\rm keV}\;,
\end{align}
where $g_\rho$ in the right-handed side of Eq.~(\ref{eq:TJscale}) should be calculated at $T_J$.
In Fig.~\ref{fig:DecayRegime}, we show   the plane of $\left(m_J, f\right)$ where the lifetime of the majoron is longer than the age of the Universe $\tau_U\simeq 4.4 \times 10^{17}$~s, and the parameter space where the majoron decays as a nonrelativistic particle (i.e., $T_J<m_J$). We also show the values of $m_J$ and $f$ that lead to the decaying temperature of $T_J=1$~eV and $T_J=0.1$~MeV,  which typically corresponds to the epochs of matter-radiation equality $T_{\rm eq}$ and the completion of the SM neutrino decoupling $T_{\nu,\rm dec}$, respectively. It can be seen from Fig.~\ref{fig:DecayRegime} that in the region of nonrelativistic majoron decay with the decaying temperature $T_{\rm eq} \lesssim T_J \lesssim T_{\nu, {\rm dec}}$, the corresponding lepton-number breaking scale resides in $f\in [1, 10^5]~{\rm GeV}$ (i.e., the yellow-shaded region in Fig.~\ref{fig:DecayRegime}), which happens to be the low-scale seesaw scenario provided that the
masses of RH neutrinos arise from the spontaneous lepton-number violation.
 
Before calculating the contribution of   majoron decay to $N_{\rm eff}$, we need to specify the early evolution of the majoron, which determines the primordial majoron abundance. Generally speaking, there are two possibilities that the majoron can keep thermal equilibrium with SM particles in the early Universe:
\begin{itemize}
\item First, the RH neutrinos can readily be thermalized in the SM plasma via two-body/inverse decay or rapid active-neutrino oscillations~\cite{Dolgov:2003sg,Li:2022bpp}. Before the electroweak gauge symmetry breaking, the majoron couples to RH neutrinos via
\begin{eqnarray}
	\label{eq:LJNNdiag}
	{\cal L}_{JNN} = -\frac{{\rm i} J}{2\sqrt{2}}\overline{N_{\rm R}^c}Y_N N_{\rm R}+{\rm h.c.} = -\frac{{\rm i} J}{2f} \sum_{i=1}^{3}M_i \overline{N_i} \gamma_5 N_i\,,
\end{eqnarray}
where the mass eigenstates of the RH neutrinos $N_{\text{R}}^{}+N_{\text{R}}^c\equiv\left(N_1,N_2,N_3\right)^{\text{T}}$ have been defined, and $M_i$ is the mass of $N_i$. Furthermore, for Yukawa couplings $Y_N$ that are not too small, the scattering $2N \to 2J$ via Eq.~(\ref{eq:LJNNdiag}) can in turn thermalize the majoron for the temperature $T \gtrsim M_i$.
Note that there is no decaying channel of $N_{i}\to N_{j}+J$ (with $M_i > M_j$) since the non-diagonal elements in $Y_{N}$ are not physical before the gauge symmetry breaking.  

\item Second, the majoron can couple with the Higgs boson via the following interaction
\begin{eqnarray}\label{eq:Higgs_portal}
	V \supset \lambda_{\Phi  S}\left(\Phi^\dagger \Phi \right)\left(S^\dagger S\right),
\end{eqnarray}
which is not forbidden by the global ${\rm U}(1)_{L}$ symmetry and should be included into a general scalar potential (see Appendix~\ref{appen:majoron-Higgs} for more details). Moreover, it was shown~\cite{Kanemura:2023jiw} that even for a portal coupling $\lambda_{\Phi S}$ as small as $\mathcal{O}(10^{-6})$, the scattering $2\Phi \to 2S$ can still thermalize the scalar $S$. After the gauge symmetry breaking, the majoron couples with the SM Higgs boson $h$ through the mixing (induced by $\lambda_{\Phi  S}$) between two CP-even bosons $h$ and $\rho$, and can be thermalized via $h\to 2J$ or $2h\to 2J$.
\end{itemize}

Therefore, a thermalized majoron in the early Universe can easily be realized in the minimal setup that exhibits a spontaneous global ${\rm U}(1)_{L}$ breaking in the scalar sector and at the same time, predicts an unstable majoron decaying at $T_{\rm eq}\lesssim T_J \lesssim T_{\nu,\rm dec}$. 
Since the majoron is much lighter than the RH neutrinos and the Higgs bosons, it is expected to undergo relativistic freeze-out.   The freeze-out temperature $T_{\rm fo}$  of the majoron depends on the details of the UV models, and will be treated as an input parameter in the following discussions. 
 
After the gauge symmetry breaking, the active neutrinos can also generate the majoron abundance through the neutrino coalescence process $2\nu\to J$ [cf. Eq.~(\ref{eq:Jnunu})]. Due to the suppression of the majoron-neutrino coupling $g_{\nu_i}\equiv m_i/f$, the production channel follows the freeze-in evolution~\cite{McDonald:2001vt,Kusenko:2006rh,Petraki:2007gq,Hall:2009bx} and culminates at $T\simeq \mathcal{O}(m_J)$. Therefore, the majoron abundance is, strictly speaking, not a constant after the relativistic freeze-out. As can be seen in Fig.~\ref{fig:DecayRegime}, for $1~\text{eV}\lesssim m_J \lesssim 1~\text{MeV}$ and a suppressed coupling $g_{\nu_i}$, the majoron is expected to decay after the SM neutrino decoupling. This late-time decay will modify the effective number of neutrino species $N_{\rm eff}$ due to the energy injection to active neutrinos. Such effects have been studied in Refs.~\cite{Chacko:2003dt,Boehm:2012gr,Escudero:2019gvw,Sandner:2023ptm,Li:2023puz} where the constraints of $N_{\rm eff}$ from the CMB measurements have been applied to derive the upper bound of the majoron-neutrino coupling in terms of the majoron mass.

Nevertheless,  previous studies have mainly focused on the majoron-neutrino effective interaction, where the majoron abundance is generated by and then depleted back to the SM neutrinos.\footnote{See, however,  Refs.~\cite{Escudero:2019gvw,Sandner:2023ptm,Li:2023puz} that partly discussed a primordial majoron abundance.} It is the purpose of this work to detail the  effects of the primordial abundance on $N_{\rm eff}$.  While we concentrate on the low-scale seesaw scenario, it is worthwhile to emphasize that the analysis performed  in subsequent sections can analogously be applied to other UV scenarios, where a new light particle coupling to active  neutrinos or photons has a non-negligible primordial abundance before the SM neutrino decoupling and decays after that epoch.

\section{Analytical derivation of $\Delta N_{\rm eff}$ from majoron decay}\label{sec:analytic_Neff}
In this section, we perform an approximate analytical calculation of the $N_{\rm eff}$  excess  from   majoron decay (i.e., $\Delta N_{\rm eff}\equiv N_{\rm eff}-N_{\rm eff}^{\rm SM}$) by assuming that the majoron decays instantaneously, where $N_{\rm eff}^{\rm SM}\simeq 3.045$ denotes the SM prediction of the effective number of neutrino species~\cite{Mangano:2001iu,Mangano:2005cc,deSalas:2016ztq,EscuderoAbenza:2020cmq,Akita:2020szl,Froustey:2020mcq,Bennett:2020zkv,Cielo:2023bqp}. The calculation depends on the kinematic properties (that is, relativistic or nonrelativistic) of the majoron at decay. In the first two subsections, we compute $\Delta N_{\rm eff}$ from relativistic and nonrelativistic majoron decay respectively, with the primordial majoron abundance    inherited from the relativistic freeze-out. In the last two subsections, we compute $\Delta N_{\rm eff}$ from the freeze-in production followed by   majoron decay, where there is no primordial majoron abundance.

\subsection{Relativistic   freeze-out and relativistic  decay}\label{sec:RfoRdec}
In order to be model-independent, we do not specify the freeze-out process of the majoron from UV scenarios (i.e., the majoron will be thermalized via the scattering either with the RH neutrinos or with the Higgs boson); rather, we treat the freeze-out temperature $T_{\rm fo}$ as an input parameter. Then the  yield of the majoron at $T_{\rm fo}$ is given by
\begin{align}
	Y^n_{J,\rm fo} & \equiv\frac{n_{J}}{s_{{\rm SM}}}\bigg |_{T=T_{\rm fo}}=\frac{45 \zeta(3)}{2\pi^4 g_{s}(T_{\rm fo})}\,,\label{eq:YJ_freeze-out}
\end{align}
where $n_{J}= \zeta(3)T^{3}/\pi^{2}$ is the number density of the relativistic majoron in thermal equilibrium with $\zeta(x)$ the Riemann $\zeta$-function, and $s_{\rm SM}$ is the SM entropy density
\begin{align}
	s_{\rm SM}=g_s(T)\frac{2\pi^2}{45}T^3\,,
\end{align} 
with $g_s(T)$ the relativistic degrees of freedom for entropy in the SM. Therefore, the majoron freeze-out yield  depends only on the relativistic degrees of freedom at  $T_{\rm fo}$. 

The observations of  the  light-element abundances generated during the BBN era put stringent constraints on the interactions of neutrinophilic particles~\cite{Huang:2017egl,Venzor:2020ova}. Here we would like to estimate the contribution of the majoron to $N_{\rm eff}$ at the BBN epoch.
It is usually stated that the BBN process starts after the deuterium bottleneck temperature $T\simeq 0.078$~MeV~\cite{Sarkar:1995dd,Cyburt:2015mya,Pitrou:2018cgg}. Nevertheless, any extra radiation before the neutron-proton freeze-out at $T\simeq 0.8$~MeV will modify the Hubble expansion rate, leading to more neutron abundance at freeze-out and hence more $^{4}\text{He}$. Observations of the primordial helium-4 synthesized at the BBN epoch will constrain the extra radiation, which is effectively parametrized by  $\Delta N_{\rm eff}$~\cite{Pitrou:2018cgg}. If the majoron decays after $T\simeq 0.1$~MeV,  the majoron itself contributes to the Hubble expansion rate  before  the BBN starts. 
Then $\Delta N_{\rm eff}$ at the BBN epoch can be approximated by\footnote{Note that the quantity $\rho_{J} /s_{\rm SM}^{4/3}$ is a constant after the majoron decouples. In addition, the majoron abundance from  the freeze-in neutrino coalescence  $2\nu\to J$ contributes negligibly to the majoron abundance and $N_{\rm eff}$ when the majoron is still in the relativistic regime.}
\begin{align}
	\Delta N_{\rm eff}^{\rm BBN}=\frac{\rho_{J}}{\rho_{\nu}^{\rm SM}}\bigg|_{T=T_{\rm BBN}}=
	\frac{\rho_J\left(T_{\rm fo}\right)}{\rho_\nu^{\rm SM}\left(T_{\rm BBN}\right)}\left[\frac{s_{\rm SM}\left(T_{\rm BBN}\right)}{s_{\rm SM}\left(T_{\rm fo}\right)}\right]^{4/3}
	=\frac{4}{7} \left(\frac{g_{s,\rm BBN}}{g_{s,\rm fo}}\right)^{4/3}, \label{eq:DeltaNeff_BBN}
\end{align}
where $g_{s,\rm BBN}$ and $g_{s,\rm fo}$ denote the relativistic degrees of freedom at $T_{\rm BBN}$ and $T_{\rm fo}$, respectively, and
\begin{align}
	\rho_\nu^{\rm SM}=\frac{7}{4} \frac{\pi^2}{30}T_\nu^4\,, 
\end{align}
is the one-flavor neutrino energy density with  $T_\nu=T_\gamma$ before neutrino decoupling. For definiteness, we will take $T_{\rm BBN}\simeq 1$~MeV to denote the temperature before the neutron-proton freeze-out.

\begin{figure}
	\centering
	
	\includegraphics[scale=0.9]{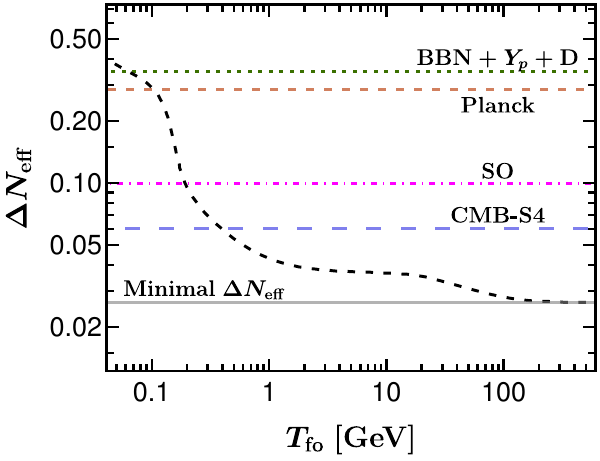} 
	
	\caption{\label{fig:Neff_Relativistic} $\Delta N_{\rm eff}$ calculated from relativistic majoron decay at the epochs of BBN or CMB with different freeze-out temperature $T_{\rm fo}$ (black short-dashed line). The regions above different horizontal lines correspond to the $2\sigma$ excluded regions of $\Delta N_{\rm eff}$ from the constraints of BBN+$Y_p$+D (green dotted line),  Planck (red short-dashed line), as well as the future projected sensitivities of SO (magenta dot-dashed line) and CMB-S4 (blue dashed line). The gray solid line denotes the minimal $\Delta N_{\rm eff}\simeq 0.027$, which corresponds to the freeze-out temperature $T_{\rm fo}\gtrsim {\cal O}(100)~{\rm GeV}$.}
\end{figure}

Next, let us consider the situation where the majoron decays relativistically after the BBN ends and prior to the matter-radiation equality epoch $T_{\rm eq}\simeq 1$~eV. The energy from the majoron would then inject into active neutrinos, acting as extra radiation at the recombination epoch $T\simeq 0.1$~eV and being constrained by the CMB measurement. In the approximation of instantaneous decay, we arrive at 
\begin{align}
\Delta N_{\rm eff}^{\rm CMB}=\frac{\rho_{J\to 2\nu}}{\rho_{\nu}^{\rm SM}}\bigg|_{T=T_{\rm eq}}=\frac{4}{7}\left(\frac{11}{4}\right)^{4/3} \left(\frac{g_{s,\rm CMB}}{g_{s,\rm fo}}\right)^{4/3},\label{eq:DeltaNeff_CMB}
\end{align}
where $\rho_{J\to 2\nu}$ denotes the neutrino energy density inherited from relativistic majoron decay, and the factor $T_\nu/T_\gamma=(4/11)^{1/3}$ has been used for the (instantaneous) neutrino decoupling.  In addition, we have assumed that the decay is completed before $T_{\rm eq}$ so that  $\Delta N_{\rm eff}$ is evaluated at $T_{\rm eq}$.

Given $g_{s,\rm BBN}/g_{s,\rm CMB}=11/4$,\footnote{This ratio is obtained in the approximation of instantaneous neutrino decoupling, which is sufficient for the analytic analysis. } we arrive at $\Delta N_{\rm eff}^{\rm BBN}=\Delta N_{\rm eff}^{\rm CMB}$. This identity implies that when the majoron decays in the relativistic regime, the extra radiation contributes equally to $N_{\rm eff}$ at the epochs of BBN and CMB, which only depends on the freeze-out temperature.  It is clear that a lower $T_{\rm fo}$ leads to a larger $\Delta N_{\rm eff}$. This can be understood as follows. After the relativistic freeze-out of the majoron, the decoupling of other SM particles will inject energies into the plasma of photons and neutrinos, thereby reheating the SM bath and enhancing $\rho_\nu^{\rm SM}$. Therefore, the later the majoron decouples, the less enhancement $\rho_\nu^{\rm SM}$ will receive, which implies a larger $\Delta N_{\rm eff}$.

In Fig.~\ref{fig:Neff_Relativistic}, we show the behavior of $\Delta N_{\rm eff}$ calculated from   relativistic majoron decay at the epochs of BBN or CMB with different freeze-out temperatures $T_{\rm fo}$, where the evolution of $g_s(T)$ is  taken from Ref.~\cite{Borsanyi:2016ksw}.  We also show the constraints on $\Delta N_{\rm eff}$ from the Planck 2018 result~\cite{Planck:2018vyg}: $	N_{\rm eff}=2.99\pm 0.17$, the combination of BBN, helium ($Y_p$) and deuterium (D) abundances~\cite{Yeh:2022heq}: $ N_{\rm eff}=2.889\pm 0.229$, and from   the future   sensitivities of Simon Observatory (SO)~\cite{SimonsObservatory:2018koc,SimonsObservatory:2019qwx} and CMB-S4~\cite{CMB-S4:2016ple,Abazajian:2019eic}. The $2\sigma$ upper bounds are given by
  \begin{align}
  \text{Planck}&: \,\Delta N_{\rm eff}<0.285\,, 
  \\[0.2cm]
    \text{BBN}+Y_p+\text{D} &: \, \Delta N_{\rm eff}<0.347\,,\\[0.2cm]
       \text{SO}&: \, \Delta N_{\rm eff}<0.1\,,\\[0.2cm]
    \text{CMB-S4}&: \, \Delta N_{\rm eff}<0.06\,.
  \end{align}
  
It can be seen that for the majoron that decays relativistically after the neutrino decoupling and before the matter-radiation equality epoch, the current constraints from BBN+$Y_p$+D and Planck requires $T_{\rm fo}>64$~MeV and $T_{\rm fo}>104$~MeV, respectively, while the future sensitivities  from the SO and CMB-S4 will further limit the decoupling temperature to $T_{\rm fo}>192$~MeV and  $T_{\rm fo}>397$~MeV, respectively.
Note that a minimal $\Delta N_{\rm eff}\simeq 0.027$ is expected for the early-time decoupling of the majoron, i.e., $T_{\rm fo}\gtrsim \mathcal{O}(100)$~GeV, where all the SM particles are relativistic with $g_{s,{\rm fo}}=106.75$. 

\subsection{Relativistic freeze-out and nonrelativistic  decay}\label{sec:RfoNRdec}
Now let us turn to  the more interesting scenario, where the majoron decays in the nonrelativistic regime. We expect that $\Delta N_{\rm eff}^{\rm BBN}\neq \Delta N_{\rm eff}^{\rm CMB}$ in this case. The former is still given by Eq.~\eqref{eq:DeltaNeff_BBN} while the latter is calculated as
\begin{align}\label{eq:DeltaNeff_RFO_NRdec}
	\Delta N_{\rm eff}^{\rm CMB}=\frac{\rho_{J\to 2\nu}}{\rho_{\nu}^{\rm SM}}\bigg|_{T=T_{\rm eq}}\simeq \frac{m_J Y^n_{J,\rm fo}s_{\text{SM} }}{\rho_{\nu}^{\rm SM}}\bigg |_{T=T_{J}}
	\simeq 0.815 \left(\frac{g_{s,T_J}}{g_{s,\rm fo}}\right)\left(\frac{m_J}{T_{J}}\right),
\end{align}
where the energy injection from the majoron into   neutrinos at $T_{\rm eq}$ is estimated by the value of $m_J n_J$ at $T_J$ [defined in Eq.~(\ref{eq:TJ})]. Note that we do not include the contribution to the majoron abundance from the freeze-in process $2\nu\to J$, so the yield $Y_J^n$ keeps constant after   freeze-out.

From Eq.~\eqref{eq:DeltaNeff_RFO_NRdec} it is clear that   later decay of the majoron will cause a larger $\Delta N_{\rm eff}$ at the CMB epoch. This can also be understood by the dilution-resistant effect~\cite{Li:2023puz} (see Sec.~\ref{subsec:compare} for more discussions): After the decoupling of neutrinos, we have $\rho_\nu^{\rm SM}\propto a^{-4}$ with $a$ the scale factor, while the majoron energy density in the nonrelativistic  regime scales as $\rho_J\propto a^{-3}$. As a result,    later decay of the majoron implies that $\rho_\nu^{\rm SM}$ suffers from more redshift (dilution) than the energy of the nonrelativistic majoron $\rho_J$ (resistance),  thereby leading to a larger $\Delta N_{\rm eff}$ after   majoron decay. 

A key observation is that, since   later decay of the majoron implies a larger lepton-number breaking scale $f$ or equivalently a smaller majoron-neutrino coupling, and leads to a larger $\Delta N_{\rm eff}^{\rm CMB}$, it implies that the constraint from $\Delta N_{\rm eff}^{\rm CMB}$ will put an upper bound on $f$. This is in contrast to the freeze-in scenario without a primordial majoron abundance, where a lower bound of $f$ can be obtained from the constraint of $\Delta N_{\rm eff}^{\rm CMB}$, as we shall discuss below.

\subsection{Freeze-in without primordial majoron}
\label{subsec:freeze-in}
In the low-scale seesaw scenario, where $f$ is around the electroweak scale, the neutrino coalescence $2\nu\to J$ can also contribute to the majoron abundance via the freeze-in mechanism. In this subsection, we calculate the excess of $N_{\rm eff}$ from freeze-in production of the majoron, and make a comparison with the results derived in the previous subsection with a primordial majoron abundance. 

To simplify the calculation of $\Delta N_{{\rm eff}}$, we assume that the freeze-in production of the majoron is essentially completed before it decays. We first consider the case of nonrelativistic majoron decay. The Boltzmann equation of the majoron abundance before decay is given by
\begin{align}\label{eq:dYJdT-fi}
	\frac{{\rm d}Y_{J,\rm fi}^n}{{\rm d}T}=-\frac{\mathcal{C}_{2\nu \to J}^{n}}{s_{\rm SM}H T}\,,
\end{align}
where $Y_{J,\rm fi}^n\equiv n_J/s_{\rm SM}$ denotes the freeze-in abundance of the majoron and the collision term $\mathcal{C}_{2\nu \to J}^{n}$ is computed in Appendix~\ref{appen:Cn_2vJ}, which is a function of the neutrino temperature $T_{\nu}$.

Let us assume again that the decay occurs instantaneously at $T_J$ after the neutrino decoupling.
Then the excess of $N_{{\rm eff}}$ at the CMB epoch coming from the nonrelativistic majoron decay can be calculated by
\begin{align}
	\Delta N_{{\rm eff}}^{{\rm CMB}} \simeq \frac{m_J Y^n_{J,\rm fi} s_{\text{SM}}}{\rho_{\nu}^{\rm SM}}\bigg |_{T=T_{J}} \simeq 2.94 g_{s,T_J}Y_{J,\rm fi}^n  \left(\frac{m_J}{T_J}\right),\label{eq:Delta_Neff_anal}
\end{align}
where $g_{s,T_J}$ denotes the relativistic degrees of freedom at the decaying temperature. Note that the nonrelativistic factor $m_J/T_J\propto \sqrt{m_J/M_{\rm Pl}}f/m_i$ [cf. Eq.~\eqref{eq:TJscale}] also appears here as in Eq.~\eqref{eq:DeltaNeff_RFO_NRdec}. 

At first sight, it seems that a heavier majoron and a larger lepton-number breaking scale lead to a larger $\Delta N_{{\rm eff}}^{{\rm CMB}}$, which occurs in the freeze-out case (see Sec.~\ref{sec:RfoNRdec}). However, $\Delta N_{\rm eff}^{\rm CMB}$ in Eq.~\eqref{eq:Delta_Neff_anal} also depends on $m_J$ and $f$ through $Y^n_{J,\rm fi}$. To see it more clearly, we calculate $Y^n_{J,\rm fi} $  with the following approximations:
\begin{itemize}
	
	\item We take the Boltzmann approximation of the neutrino distribution functions so that the collision term $C_{2\nu \to J}^n$ is given by Eq.~\eqref{eq:Cn_nunuJ_Boltzmann}.
	
	\item We assume that neutrinos decouple instantaneously at $T_{\nu,\rm dec}\simeq 0.1$~MeV, so the neutrino temperature in $C_{2\nu \to J}^n$ is given by $T_\nu = T$ (for $T>T_{\nu,\rm dec}$) or $T_\nu/T=(4/11)^{1/3}$ (for $T<T_{\nu,\rm dec}$).
	
	\item To obtain the freeze-in abundance of the majoron, one needs to integrate Eq.~(\ref{eq:Delta_Neff_anal}) over the temperature. Since the majoron decays nonrelativistically after the neutrino decoupling, we can further expand the collision term in the integral in terms of $m_J/T$ for $T>T_{\nu,\rm dec}$ and in terms of $T/m_J$ for $T<T_{\nu,\rm dec}$.
\end{itemize}
With above approximations, we arrive at a very simple result 
\begin{align}\label{eq:DeltaNeff_fi}
\Delta N_{{\rm eff}}^{{\rm CMB}} \simeq 0.139 \left(\frac{\text{MeV}}{m_J}\right)^{1/2}\left(\frac{\rm GeV}{f}\right),
\end{align}
where $g_{s,T_J}\simeq 3.36$ has been used, and for concreteness, we have taken the normal ordering of the neutrino masses with $m_1=0$ and $\sum_{i=1}^{3} m_i^2 \simeq \left(0.0086~{\rm eV}\right)^2+\left(0.05~{\rm eV}\right)^2$~\cite{Esteban:2020cvm}. 

From Eq.~(\ref{eq:DeltaNeff_fi}) it is now clear that for the freeze-in case, a lighter majoron and a smaller $f$ will lead to a larger $\Delta N_{{\rm eff}}$. Therefore,  the constraint from CMB measurements will put a lower bound on the lepton-number breaking scale. This is in contrast to the case of freeze-out with a primordial majoron abundance, as already discussed in Sec.~\ref{sec:RfoNRdec}.


\subsection{Comparison between freeze-out and freeze-in}
\label{subsec:compare}
From Eq.~(\ref{eq:DeltaNeff_fi}), one may naively expect that $\Delta N_{\rm eff}$ can keep increasing with the decreasing of $m_J$ and $f$. However, a lighter majoron and a larger coupling (i.e., a smaller $f$) would make the majoron-neutrino system easier to reach thermal equilibrium, when the freeze-in formalism and consequently the result in Eq.~(\ref{eq:DeltaNeff_fi}) are no longer applicable. In fact, as pointed out in Refs.~\cite{EscuderoAbenza:2020cmq,Li:2023puz}, when the majoron gets thermalized with the SM neutrinos, one obtains a maximal $\Delta N_{\rm eff} \simeq 0.12$.  
  
Based on the analysis in Sec.~\ref{subsec:freeze-in}, we can understand the maximal $\Delta N_{\rm eff}$ in a different way.\footnote{In Refs.~\cite{EscuderoAbenza:2020cmq,Li:2023puz}, the maximal $\Delta N_{\rm eff}$ is derived from the conservation of particle number, energy and entropy. }  Recall that the result in Eq.~\eqref{eq:DeltaNeff_fi} is derived in the regime of nonrelativistic majoron decay, i.e.,  $p_J<m_J$  with $p_J$   the magnitude of the majoron momentum. The averaged momentum for a nonrelativistic thermalized majoron is given by $\langle p_J \rangle \simeq \sqrt{3m_J T_J}$.\footnote{Strictly speaking, the majoron never reaches thermal equilibrium in our analysis of the freeze-in process. Nevertheless, here we use the statistics for a thermalized majoron since we only want to estimate the upper bound of $\Delta N_{\rm eff}$ that is caused by the majoron decay.} By requiring $\langle p_J \rangle < m_J$ and taking advantage of Eq.~(\ref{eq:TJscale}), we obtain a lower bound on $f$: 
\begin{align}
	\label{eq:NRcondition}
	\left(\frac{f}{\text{GeV}}\right) \gtrsim 0.96 \left(\frac{\text{MeV}}{m_J}\right)^{1/2},
\end{align}
where $\sum_{i=1}^{3} m_i^2 \simeq \left(0.0086~{\rm eV}\right)^2+\left(0.05~{\rm eV}\right)^2$ and $g_\rho(T_J)\simeq 3.36$ have been used.
Substituting Eq.~(\ref{eq:NRcondition}) back to Eq.~\eqref{eq:DeltaNeff_fi}, we arrive at $\Delta N_{\rm eff}^{\rm CMB} \lesssim 0.14$. This confirms the result in Refs.~\cite{EscuderoAbenza:2020cmq,Li:2023puz} that   nonrelativistic majoron decay from the freeze-in mechanism without a primordial abundance can lead to a maximal $\Delta N_{\rm eff}$ at ${\cal O} (0.1)$.   
  
Another point worth mentioning is that there is no excess of $N_{\rm eff}$ from the relativistic majoron decay in the freeze-in scenario. In Sec.~\ref{sec:RfoRdec}, we have demonstrated that the relativistic majoron decay with a primordial abundance can generate a nonzero $\Delta N_{\rm eff}$. However, this is not the case if the majoron abundance only comes from the freeze-in production, which instead predicts a vanishing $\Delta N_{\rm eff}$.  To see it more clearly, we can rewrite the Boltzmann equation of the majoron-neutrino system  as~\cite{Li:2023puz}
 \begin{align}
 	\frac{{\rm d}}{{\rm d}a} \left[(\rho_J+\rho_\nu)a^4\right]=a^3 (\rho_J-3P_J)\,,
 \end{align}
where $P_J$ denotes the pressure of the majoron. In the relativistic regime, we have $P_J=\rho_J/3$, so the comoving energy density $\left(\rho_J+\rho_\nu\right) a^{4}$ is a constant. In this case, if there is no primordial majoron abundance,  the energy density will first transfer from neutrinos to the majoron through the freeze-in process $2\nu\to J$ and then back to   neutrinos  through the relativistic majoron decay $J \to 2\nu$, while the total energy density of the majoron-neutrino system in a comoving volume remains unchanged. Therefore, there would be no excess of $N_{\rm eff}$ if the majoron decays in the relativistic regime.
  
In conclusion,  if there is no primordial majoron abundance, a nonzero $\Delta N_{\rm eff}$ can only be generated if the majoron decays in the nonrelativistic regime. In addition,  a smaller lepton-number breaking scale $f$ and majoron mass $m_J$ lead to a larger $\Delta N_{\rm eff}$, which has a maximal value at  ${\cal O} (0.1)$. On the other hand, when there is a primordial majoron abundance, both the relativistic and nonrelativistic majoron decay at late times can generate a nonzero $\Delta N_{\rm eff}$. In particular, for the case of nonrelativistic decay, a larger $f$ and $m_J$ will predict a larger $\Delta N_{\rm eff}$, contrary to the freeze-in case where there is no primordial abundance. 
To have a more precise constraint on the lepton-number breaking scale from observations of $N_{\rm eff}$, in the next section, we proceed to perform a stricter calculation of $\Delta N_{\rm eff}$ in the nonrelativistic region by going beyond the approximation of   instantaneous majoron decay  used in Sec.~\ref{sec:RfoNRdec}.
  
\section{Calculation of $\Delta N_{\rm eff}$ beyond instantaneous majoron decay}\label{sec:numerics}
In this section, we carry out a stricter calculation of $\Delta N_{\rm eff}$ from   nonrelativistic majoron decay, which has  a primordial abundance inherited from the relativistic freeze-out. We also make a comparison with   the freeze-in production $2\nu\to J$ followed by   nonrelativistic decay $J\to 2\nu$. For later reference, we use the following shorthand:
\begin{align*}
\text{FONR}&\equiv	\text{relativistic~Freeze-Out}+\text{NonRelativistic decay}\,,
\\[0.2cm]
\text{FINR}&\equiv	\text{Freeze-In}+\text{NonRelativistic decay}\,.
\end{align*}

We are interested in the case where the majoron decays after the SM neutrinos have
decoupled from the plasma at around $T_{\nu,\rm dec}\simeq 0.1$~MeV~\cite{deSalas:2016ztq,Gariazzo:2019gyi,Froustey:2020mcq,EscuderoAbenza:2020cmq,Akita:2020szl,Bennett:2020zkv,Cielo:2023bqp}, so as to suppress the nontrivial impacts  on the BBN processes as well as the neutrino decoupling. Under this circumstance, the neutrinos generated from   majoron decay at temperatures below $T_{\nu,\rm dec}$ can no longer be thermalized via the SM weak interactions. Therefore, in the FONR case, we can treat   majoron decay separately from the SM thermal bath. 

In general, the neutrino coalescence $2\nu\to J$ also contributes to $\Delta N_{\rm eff}$ in the FONR case. Nevertheless, we will neglect it in the calculation of $\Delta N_{\rm eff}$  for two reasons. First, as shown in Sec.~\ref{subsec:compare}, the contribution from neutrino coalescence to $\Delta N_{\rm eff}$ can only reach up to ${\cal O} (0.1)$, which is beyond the sensitivity of current CMB measurements~\cite{Planck:2018vyg}. Second, as can be seen from Sec.~\ref{sec:RfoNRdec} and Sec.~\ref{subsec:freeze-in}, the dependence of $\Delta N_{\rm eff}$ on $m_J$ and $f$ is opposite between the FONR and the FINR cases, so it would be more clear to treat the two  processes separately.  The inclusion of neutrino coalescence in the FONR case will be discussed in Sec.~\ref{sec:constraints}.

Without including the contribution from neutrino coalescence,  the Boltzmann equations governing the nonrelativistic majoron decay are given by
\begin{align}\label{eq:drhodt_fo}
	\frac{{\rm d}\rho_{J}}{{\rm d}t}+3H\rho_{J} =-\mathcal{C}_{J\to2\nu}^{\rho}\,, \quad	\frac{{\rm d}\rho_{\nu}}{{\rm d}t}+4H\rho_{\nu}  =\mathcal{C}_{J\to2\nu}^{\rho}\,,
\end{align}
where $\mathcal{C}_{J\to2\nu}^{\rho}$ is the collision term responsible for the energy transfer rate. In the nonrelativistic regime, we have $\mathcal{C}_{J\to 2\nu}^{\rho}\simeq m_{J}n_{J}\Gamma_{J}$ with $\Gamma_{J}\simeq \Gamma_{J\to 2\nu}$ and the primordial majoron abundance $n_J =\zeta(3)T^3/\pi^2$ coming from relativistic freeze-out.

Defining  $Y_{J}^n\equiv n_J/s_{\rm SM} = \rho_{J}/\left(m_J s_{\rm SM}\right)$ and  $Y_{\nu}^\rho\equiv \rho_\nu/s_{\rm SM}^{4/3}$, we can rewrite the Boltzmann equations as
\begin{align}\label{eq:dYdt_fi}
	\frac{{\rm d}Y_{J}^n}{{\rm d}T}=\frac{\Gamma_{J}Y_{J}^n}{HT}\,,\qquad
	\frac{{\rm d}Y_{\nu}^{\rm \rho}}{{\rm d}T}=-\frac{m_J \Gamma_{J} Y_{J}^n}{s_{\rm SM}^{1/3}HT}\,,
\end{align}
which lead to the solutions
\begin{align}\label{eq:YJYnu}
	Y_{J}^n(T)= Y_{J,\rm ini}^n\, e^{-\frac{\Gamma_{J}}{2}\left(H^{-1}-H_{\rm ini}^{-1}\right)}\,,\qquad
	Y_{\nu}^{\rho}(T)=m_{J}\Gamma_{J}\int_{T}^{T_{\rm ini}}{\rm d}T' \frac{Y_{J}^n(T')}{s_{{\rm SM}}^{1/3}HT'} \,,
\end{align}
with an initial temperature  $T_{\rm ini}\gg T_{J}$ and $H_{\rm ini}\equiv H(T_{\rm ini})$. For definiteness,  we take $T_{\rm ini}=0.1$~MeV, which is consistent with the requirement of $T_J \lesssim T_{\nu,\rm dec}$. It should be pointed out that a higher $T_{\rm ini}$ does not modify the result significantly, since the decay mainly occurs around $T_J$ determined by $\Gamma_{J}\simeq 2H (T_J)$, and $Y_J^n$ is exponentially suppressed for $T < T_J$.\footnote{This exponential suppression also justifies the approximate analytical results derived in Sec.~\ref{sec:analytic_Neff} under the approximation of instantaneous majoron decay.}  In addition, the initial abundance is determined by the freeze-out abundance in Eq.~\eqref{eq:YJ_freeze-out}, i.e., 
\begin{align}
	\label{eq:initial}
	Y_{J,\rm ini}^n= Y_{J,\rm fo}^n = \frac{45 \zeta(3)}{2\pi^4 g_{s}(T_{\rm fo})}\,.
\end{align}

\begin{table}[t!]
	\normalsize 
	\setlength{\tabcolsep}{6pt}
	\renewcommand{\arraystretch}{1.2}
	\centering
	\begin{tabular}{c|c|c|c|c|c|c|c}
		\hline \hline
		$T_{\rm fo}$/MeV & 64 & 104 & 192 & 397 & $>{\cal O} (10^5)$ &---& ---\\
		\hline
		$Y_{J,{\rm ini}}^n$ & 0.018 & 0.015 & 0.007 & 0.005 & 0.003 &  $9 \times 10^{-4}$  & $3 \times 10^{-4}$  \\
		\hline
		$\Delta N_{\rm eff}^{\rm BBN}$ & 0.347 & 0.285 & 0.100 & 0.060 &  0.027 & 0.008 & 0.003\\
		\hline\hline
	\end{tabular}
	\caption{\label{table:correspondence} Correspondence among the freeze-out temperatures $T_{\rm fo}$, the primordial majoron abundances $Y_{J,{\rm ini}}^n$, and the contributions to $\Delta N_{\rm eff}$ at the BBN epoch.}
\end{table}

In Tab.~\ref{table:correspondence}, we have listed the correspondence between $T_{\rm fo}$ and $Y_{J,\rm ini}^n$ for some typical freeze-out temperatures. Alternatively, one can also use the contribution of the relativistic majoron to $\Delta N_{\rm eff}$  at the BBN epoch to characterize the primordial abundance (i.e., the third line of Tab.~\ref{table:correspondence}). Note that the first five values of $\Delta N_{\rm eff}^{\rm BBN}$ in Tab.~\ref{table:correspondence} can be obtained from $T_{\rm fo}$ using Eq.~(\ref{eq:DeltaNeff_BBN}). For $\Delta N_{\rm eff}^{\rm BBN}<0.027$, it cannot be inherited from the relativistic freeze-out mechanism,  where the thermal plasma at $T>\mathcal{O}(100)$~GeV only contains the SM degrees of freedom.  Nevertheless, we can still establish a one-to-one  correspondence between $\Delta N_{\rm eff}^{\rm BBN}$ and the primordial majoron abundance $Y_{J,{\rm ini}}^n$.

Since the nonrelativistic majoron decay occurs after the neutrino decoupling and before the matter-radiation equality epoch, when the relativistic species include photons and neutrinos, it is reasonable to treat the relativistic degrees of freedom $g_\rho$ (in the Hubble parameter) and $g_s$ (in the entropy density) as constants. Taking into account the temperature ratio between photons and neutrinos derived from noninstantaneous neutrino decoupling: $T/T_\nu \simeq 1.3985$~\cite{EscuderoAbenza:2020cmq}, one obtains
\begin{align}
	\label{eq:DOF1}
g_\rho &\simeq 2+\frac{7}{8}\times 2\times3\times \left(\frac{T_\nu}{T}\right)^{4}\simeq 3.38\,,\\
g_s    &\simeq 2+\frac{7}{8}\times 2\times3\times \left(\frac{T_\nu}{T}\right)^{3}\simeq 3.93\,. 	\label{eq:DOF2}
\end{align}

Combining Eqs.~(\ref{eq:YJYnu})-(\ref{eq:DOF2}) and given a freeze-out temperature $T_{\rm fo}$, one can obtain the neutrino energy density $Y_\nu^\rho$ that comes from the majoron decay. The excess of $N_{\rm eff}$ at the CMB epoch is then given by
\begin{align}
	\Delta N_{{\rm eff}}^{{\rm CMB}}=\frac{Y_{\nu}^\rho  s_{\rm SM}^{4/3}}{\rho^{\rm SM}_{\nu}}\bigg |_{T=T_{\rm eq}}.\label{eq:Delta_Neff_NRFO}
\end{align}

Note that $\Delta N_{\rm eff}$ 
calculated above is based on Eq.~\eqref{eq:drhodt_fo}, where only the contribution from majoron decay $J\to 2\nu$ is included.  On the other hand, the heavy sterile neutrinos $N_i$ could also contribute to $\Delta N_{\rm eff}$ through decay ($N_i \to J + \nu_j$) or 2-to-2 scattering ($2J \to 2\nu$ mediated by $N_i$). In the following, we show that the contributions from sterile neutrinos to $\Delta N_{\rm eff}$ are negligible compared with that from majoron decay.
\begin{itemize}
    \item We first consider the contribution from sterile neutrino decay $N_i\to J + \nu_j$ (for $i,j=1,2,3$). Since the masses of sterile neutrinos satisfy $M_i\sim f\gg T_{\nu,{\rm dec}}$, the decay occurs much earlier than the decoupling of active neutrinos. So the direct contribution from decay to $\Delta N_{\rm eff}$ is washed out as active neutrinos are still in thermal equilibrium with photons. On the other hand,  the freeze-in production of majoron abundance from $N_i$ decay culminates at $T\simeq \mathcal{O}(M_i)$, at which the relativistic majoron freeze-out also takes place. At this epoch, the contribution to primordial majoron abundance is dominated by the thermal freeze-out, while the freeze-in contribution is negligible.
    
    \item Next we consider the contribution from $2J \to 2\nu$ mediated by $N_i$ (for $i=1,2,3$). We can compute the cross section of $2J\to 2\nu$ using Eq.~\eqref{eq:JNnu}, with the coupling suppressed by ${\cal C}_{ij}\sim Y_\nu \lesssim 10^{-5}$ in the low-scale seesaw scenario. The cross section is $s$-wave dominated, so that we can estimate the collision term as follow:
\begin{align}
	\mathcal{C}^\rho_{2J\to 2\nu}\sim \langle\sigma v\rangle_{2J\to2\nu}n_J^2 m_J\sim  \frac{m_J n_J^2}{ f^4}\sum_{i=1}^{3}m_{i}^{2}\,.
	\end{align} 
Here we have neglected the exact numerical factor from phase-space integration, which is smaller than that from majoron decay. Recall that the collison term from majoron decay is given by ${\cal C}_{J\to 2\nu}^\rho = m_J n_J \Gamma_J$. So we obtain $\mathcal{C}^\rho_{J\to 2\nu}/\mathcal{C}^\rho_{2J\to 2\nu} \sim m_J f^2/T^3$. Given $m_J\gtrsim T$ for nonrelativistic majoron decay/annihilation and $f\gg m_J$, we conclude that the contribution to $\Delta N_{\rm eff}$ from majoron annihilation $2J\to 2\nu$ is strongly suppressed compared with that from majoron decay $J\to 2\nu$. 
\end{itemize}


\begin{figure}
	\centering
	
	\includegraphics[scale=0.95]{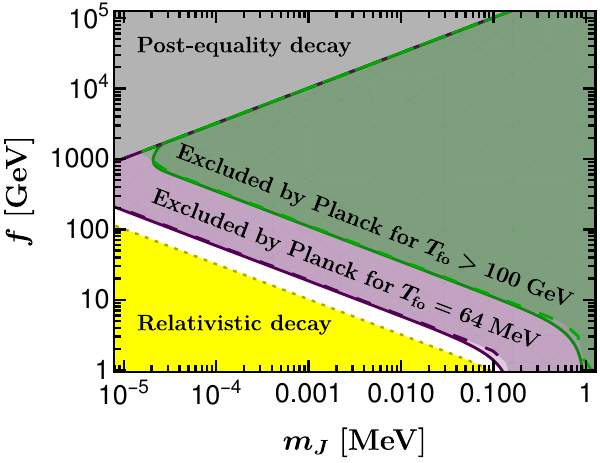} 
	
	\caption{\label{fig:Neff_CMB} Allowed parameter space of the majoron mass $m_J$ and the lepton-number breaking scale $f$ under the constraints of current CMB measurements on $N_{\rm eff}$. The pink- and green-shaded regions in the upper right corner are excluded by Planck 2018 at 2$\sigma$ level~\cite{Planck:2018vyg}, with different primordial majoron abundances characterized by the freeze-out temperatures $T_{\rm fo}$. The excluded regions are obtained from  instantaneous decay  in Sec.~\ref{sec:RfoNRdec} (dashed line) and noninstantaneous decay in Sec.~\ref{sec:numerics} (solid line), respectively. In addition, the yellow-shaded region in the lower left corner corresponds to the relativistic decay regime, while the gray-shaded region in the upper left corner corresponds to the scenario where the majoron decays after the matter-radiation equality epoch (post-equality decay).}
\end{figure}

Based on our calculations of $\Delta N_{\rm eff}$ discussed above, the excluded regions in the $(m_J, f)$ plane from the current CMB measurements 
are shown in Fig.~\ref{fig:Neff_CMB}. To compare the calculation of $\Delta N_{\rm eff}$ using the approximation of instantaneous majoron decay in Sec.~\ref{sec:RfoNRdec} with   non-instantaneous decay in Sec.~\ref{sec:numerics}, we have shown the excluded regions obtained from both in Fig.~\ref{fig:Neff_CMB}, which match quite well with each other. Therefore, the instantaneous majoron decay serves as a good approximation in our problem.

The excess of $N_{\rm eff}$ from the majoron decay depends on the primordial majoron abundance, which is characterized by the freeze-out temperature $T_{\rm fo}$ in Eq.~(\ref{eq:initial}). In Fig.~\ref{fig:Neff_CMB}, we have shown results from two typical freeze-out temperatures: $T_{\rm fo} = 64~{\rm MeV}$ and $T_{\rm fo} > 100~{\rm GeV}$. The first one is the lowest freeze-out temperature that is allowed by BBN+$Y_p$+D (see Fig.~\ref{fig:Neff_Relativistic}). A lower $T_{\rm fo}$ corresponds to a larger primordial abundance and would lead to a larger $\Delta N_{\rm eff}$ excluded by the BBN measurements [cf. Eq.~(\ref{eq:DeltaNeff_BBN})]. The second one is the case where the freeze-out temperature is sufficiently high and all the SM species are relativistic, corresponding to a minimal $\Delta N_{\rm eff}\simeq 0.027$ at the BBN epoch. We can see that for the largest primordial abundance allowed by current BBN measurements, the majoron that decays nonrelativistically is severely constrained. In particular, a lepton-number breaking scale above 300~GeV is excluded when the majoron mass is within $[10^{-5},1]$~MeV. The constraint is less strict with a smaller primordial abundance or equivalently, a higher freeze-out temperature. For $T_{\rm fo} > 100~{\rm GeV}$, which corresponds to the smallest primordial abundance that can be induced from relativistic freeze-out, we find that a breaking scale $f>2~{\rm TeV}$ is already excluded by the current Planck measurements  for $m_J \in [10^{-5},1]$~MeV.  

Note that for $f<1$~GeV, the majoron decay generally occurs in the relativistic regime, as can be inferred from  Fig.~\ref{fig:Neff_CMB}. Therefore, we   concentrate on $f>1$~GeV. Also, it is worthwhile to mention that for a lighter majoron around the eV scale, the nonrelativistic decay could occur after the epochs of matter-radiation equality and recombination when effects on, e.g., the structure formation and the neutrino free-streaming become significant~\cite{Sandner:2023ptm}. On the other hand, for a heavier majoron above the MeV scale, the decay channel to electron-positron pairs opens. Besides, the majoron decay to neutrinos and electron-positron pairs could occur during the epochs of the BBN and neutrino decoupling and have observable effects. In particular, for $1~{\rm MeV} \lesssim m_J \lesssim 100~{\rm MeV}$, the most stringent constraint on $f$ comes from Supernova 1987A. The absence of observing $100~{\rm MeV}$-range neutrino events from Supernova 1987A puts a lower bound on the lepton-number breaking scale: $f \gtrsim 0.1~{\rm GeV} \left(m_J/{\rm MeV}\right)$~\cite{Fiorillo:2022cdq}. See also~\cite{Akita:2022etk,Akita:2023iwq} for further discussions. It would be interesting to further investigate these nontrivial effects outside the mass region shown in Fig.~\ref{fig:Neff_CMB}, which goes beyond the scope of the current work.


\section{The sandwiched  window from precision $N_{\rm eff}$ measurements}
\label{sec:constraints}
We have seen from Fig.~\ref{fig:Neff_CMB} that the current Planck measurements of $N_{\rm eff}$ can already put strict upper bounds on the lepton-number breaking scale for the scenario where the majoron decays nonrelativistically with a primordial abundance (FONR). On the other hand, in Sec.~\ref{subsec:freeze-in} we demonstrated that the majoron abundance can also be accumulated through the freeze-in production $2\nu\to J$ and later decays nonrelativistically back to the neutrinos (FINR). As discussed in Sec.~\ref{subsec:compare}, the FINR case can lead to a maximal $\Delta N_{\rm eff}\sim {\cal O} (0.1)$, which is beyond the sensitivity of the current Planck measurements. 

However, the forecast sensitives of $\Delta N_{\rm eff}$ measurements in the future CMB experiments will be increased by a factor of a few. For instance, the SO experiment~\cite{SimonsObservatory:2018koc,SimonsObservatory:2019qwx} has a projected $2\sigma$ sensitivity $\Delta N_{\rm eff}<0.1$,  while the CMB-S4 experiment~\cite{CMB-S4:2016ple,Abazajian:2019eic} is expected to have a $2\sigma$ sensitivity $\Delta N_{\rm eff}<0.06$. Therefore, for future CMB experiments, the FINR case can also be probed. Furthermore, due to the opposite dependence  of $\Delta N_{\rm eff}$ on $m_J$ and $f$ between the FONR and the FINR cases, one can expect that the viable parameter space of the lepton-number breaking scale would be pushed into a narrow sandwiched window when   the contributions from the FONR and FINR cases coexist. In particular, null signals of the  $N_{\rm eff}$ excess in future CMB experiments could even close such a window and completely exclude the model.

\begin{figure}
	\centering
	
	\includegraphics[scale=0.67]{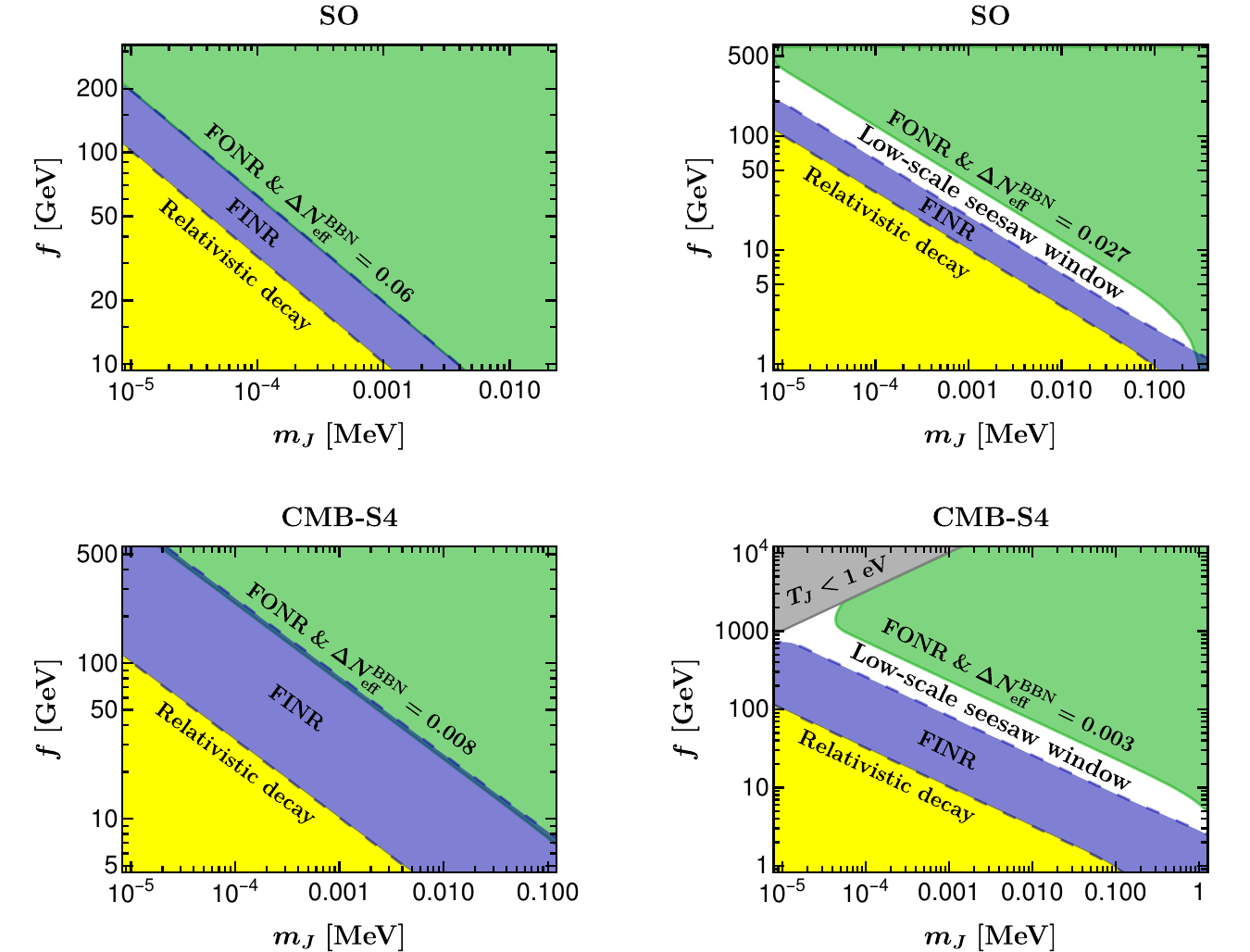}
	\caption{\label{fig:Neff_SO_CMBS4} The sandwiched window for the lepton-number breaking scale from the constraints of the future SO (upper panel: $\Delta N_{\rm eff}<0.1$) and CMB-S4 (lower panel: $\Delta N_{\rm eff}<0.06$) experiments. In each subfigure, the green-shaded regions denote the excluded regions from the FONR case with different primordial majoron abundances (characterized by $\Delta N_{\rm eff}^{\rm BBN}$). The shaded regions in the lower left corner denote the regime of relativistic decay (yellow) and the regions excluded from the FINR cases (purple). The white bands labeled by ``low-scale seesaw window'' are the viable parameter space where the contributions from FONR and FINR cases coexist.}  
\end{figure}

To see the sandwiched window more clearly,  we first consider the bounds from the FONR and FINR cases separately. That is, the bound from one case is derived without the contribution from the other case. In Fig.~\ref{fig:Neff_SO_CMBS4}, we show the sandwiched window for the lepton-number breaking scale from the constraints of future SO and CMB-S4 experiments at 2$\sigma$ level. Note that for the FONR case, the constraints depend on the primordial majoron abundances, which are characterized by $\Delta N_{\rm eff}^{\rm BBN}$, i.e., the contributions of the primordial relativistic majoron  at the BBN epoch (see Tab.~\ref{table:correspondence}). For the FINR case, we apply the results derived in Ref.~\cite{Li:2023puz} by replacing the effective majoron-neutrino coupling $g_\nu$ with the breaking scale via $f=0.05~\text{eV}/g_\nu$.  It corresponds to mapping  the model-independent results into the seesaw framework.

In the upper left panel of Fig.~\ref{fig:Neff_SO_CMBS4}, we show the parameter space of $m_J$ and $f$ under the constraints of SO with a primordial majoron abundance characterized by $\Delta N_{\rm eff}^{\rm BBN}= 0.06$ (corresponding to $T_{\rm fo}=397~{\rm MeV}$). It can be seen that for such a primordial abundance, the excluded region from the FONR case happens to be adjacent to that from the FINR case, and there is no viable parameter space for   nonrelativistic majoron decay. This implies that if an excess of $\Delta N_{\rm eff}^{\rm CMB} > 0.1$ is not observed by the future SO experiment at 2$\sigma$ level, then the scenario with the freeze-out temperature $T_{\rm fo} < 397$~MeV will be ruled out within the regime of nonrelativistic decay.  

In the upper right panel of Fig.~\ref{fig:Neff_SO_CMBS4}, the result is shown for a smaller primordial majoron abundance characterized by $\Delta N_{\rm eff}^{\rm BBN}= 0.027$ (corresponding to $T_{\rm fo}>100~{\rm GeV}$). Since the primordial abundance is reduced, the excluded region from the FONR case is expected to be smaller than that in the upper left panel, thereby leaving a viable parameter space that is sandwiched by the constraints from the FONR and the FINR cases (denoted as  ``low-scale seesaw window''). For such a low-scale seesaw scenario, this is the  parameter space  where a minimal primordial majoron abundance inherited from  relativistic freeze-out can survive, provided that no excess of $\Delta N_{\rm eff}^{\rm CMB}\geqslant 0.1$ is observed by the future SO  experiment. 


In the lower left and lower right panels of Fig.~\ref{fig:Neff_SO_CMBS4}, the results are shown under the constraints of CMB-S4 with a primordial majoron abundance characterized by $\Delta N_{\rm eff}^{\rm BBN}= 0.008$ and $\Delta N_{\rm eff}^{\rm BBN}= 0.003$, respectively.\footnote{Note that for $\Delta N_{\rm eff}^{\rm BBN}< 0.027$, the primordial majoron abundance cannot be inherited from the relativistic freeze-out (see Fig.~\ref{fig:Neff_Relativistic}). Smaller primordial abundances may be satisfied by other mechanisms (e.g., freeze-in production from the Higgs boson or RH neutrinos in the early Universe) or by the existence of extra relativistic degrees of freedom at high temperatures.} We can infer from the lower left panel that the future CMB-S4 experiment is able to fully close the window for the low-scale seesaw scenario, where the primordial majoron abundance comes from relativistic freeze-out and later decays nonrelativistically, provided that an excess of $\Delta N_{\rm eff}^{\rm CMB}\geqslant 0.06$ is not observed by CMB-S4 at 2$\sigma$ level. Nevertheless, the case with a smaller primordial abundance ($\Delta N_{\rm eff}^{\rm BBN}< 0.008$) is still possible to survive if the abundance comes from other mechanism rather than relativistic freeze-out, as is shown in the lower right panel. This implies that even if the majoron itself as a relativistic species at the BBN epoch only contributes to $\Delta N_{\rm eff}$ at the order of $0.1\%$, its late-time nonrelativistic decay can   generate observable effects in the future CMB experiments.

Finally, it should be noted that the above analysis is based on the assumptions that the bounds from the FONR and FINR cases are treated separately. For the real situation, the contribution  to $\Delta N_{\rm eff}$ from neutrino coalescence $2\nu\to J$ should also be taken into account in the FONR case. In this case, one expects that the sandwiched window in Fig.~\ref{fig:Neff_SO_CMBS4} will become somewhat  narrower. This can be understood by the fact that the contribution to $\Delta N_{\rm eff}$ at the CMB epoch for the real situation can always be divided into two parts
\begin{align}
	\Delta N_{\rm eff}^{\rm CMB} = \Delta N_{\rm eff}^{\rm FO} + \Delta N_{\rm eff} ^{\rm FI}\;,
\end{align}
where $\Delta N_{\rm eff}^{\rm FO}$ ($\Delta N_{\rm eff}^{\rm FI}$) denotes the contribution from the primordial majoron abundance (neutrino coalescence). Since both $\Delta N_{\rm eff}^{\rm FO}$ and $\Delta N_{\rm eff}^{\rm FI}$ are \emph{non-negative}, one will always obtain a larger $\Delta N_{\rm eff}$ than taking into account $\Delta N_{\rm eff}^{\rm FO}$ and $\Delta N_{\rm eff}^{\rm FI}$ separately. Therefore, the constraints and exclusion regions that we obtained in Fig.~\ref{fig:Neff_SO_CMBS4} should be considered as the most conservative results. Nevertheless, since $\Delta N_{\rm eff}^{\rm FI}$ can reach a maximal value at the order of 0.1 and the dependence of $\Delta N_{\rm eff}^{\rm FI}$ on $m_J$ and $f$ is opposite to $\Delta N_{\rm eff}^{\rm FO}$, it justifies our treatment as a good approximation to visualize
the sandwiched window. 

Before closing this section, we briefly comment on the applicability of our results to other well-motivated scenarios that feature lepton-number violations.  In the low-scale type-I seesaw framework concerned here, the Dirac Yukawa coupling is highly suppressed by the active neutrino mass, i.e., $Y_\nu\sim \sqrt{m_i f}/v \lesssim 10^{-5}$. Such small couplings could be evaded, e.g., in the framework of the inverse seesaw model~\cite{Mohapatra:1986bd}. In the inverse seesaw scheme, apart from three RH neutrinos $N_{\rm R}$, one introduces another three singlets $N_{\rm L}$. The Lagrangian is given by
\begin{align}
{\cal L}= -\overline{\ell_{\rm L}}Y_\nu \widetilde{\Phi} N_{\rm R}-\overline{N_{\rm L}}M N_{\rm R}-\frac{1}{2} \overline{N_{\rm L}}Y_N' N_{\rm L}^c S + {\rm h.c.}\,,
\end{align}
where $M$ is the mass scale of heavy sterile neutrinos. The Majorana mass $\mu=Y_N' f/\sqrt{2}$ explicitly breaks lepton-number symmetry and thus can be very small in a technically natural way~\cite{tHooft:1979rat}. Assuming the hierarchy $\mu \ll Y_\nu v \ll M$, the active neutrino mass is given by: $m_\nu \sim Y_\nu^2 v^2 \mu/M^2$, where we have suppressed the flavor indices for simplicity. Therefore, one can have an ${\cal O}(1)$ Yukawa coupling $Y_\nu$ in the low-scale inverse seesaw scenario since the active neutrino mass is naturally suppressed by small $\mu$. In this case, we find the coupling between majoron and active neutrinos turns out to be $g_\nu \sim \sqrt{Y_N' m_\nu/f}$. Furthermore, if the lepton-number breaking scale $f$ is around the electroweak scale, then we have $g_\nu \sim m_\nu/f$, which is the same order as the case of the type-I seesaw model [see Eq.~(\ref{eq:Jnunu})]. Therefore, our above calculation of $\Delta N_{\rm eff}$ from majoron decay is also applicable to the inverse seesaw scheme. 

While the  majoron cannot serve as dark matter   in  the low-scale type-I seesaw scenario, the dark matter candidate can be readily accommodated in the model-dependent setup. In particular, under the inverse seesaw framework, some of the sterile neutrinos could be at the keV-MeV scale and play the role of warm dark matter~\cite{Fernandez-Martinez:2021ypo}. For such warm dark matter, its  contribution to $\Delta N_{\rm eff}$ is negligible due to the Lyman-$\alpha$
forest constraints~\cite{Li:2021okx}.

\section{Conclusions}
\label{sec:conclusion}
In this work, we have focused on the low-scale seesaw scenario associated with a singlet majoron from the spontaneous global ${\rm U}(1)_L$ breaking, and investigated the   constraints on the lepton-number breaking scale $f$ from the cosmological measurements of  $N_{\rm eff}$. This provides a complementary approach
to the collider searches for lepton-number violation and TeV-scale Majorana neutrinos.

For the low-scale seesaw scenario, the breaking scale $f$ is expected to be not far above the electroweak scale, so the majoron can decay within the cosmological time scale. 
The majoron interactions in the early Universe have two possible effects on $N_{\rm eff}$ at the epoch of CMB. The first is that the majoron abundance is accumulated through   freeze-in production ($2\nu\to J$) after the electroweak gauge symmetry breaking, and later decays nonrelativistically back to   neutrinos   ($J\to 2\nu$). This possibility has been widely discussed in the literature, where   lower bounds of $f$ were obtained.
In this work, we mainly focused on the second possibility that was usually neglected in previous studies. That is, the majoron possesses a non-negligible primordial abundance, which is inherited from relativistic freeze-out and later depleted  into neutrinos. This situation is quite common within the minimal framework, where the majoron gets thermalized in the SM bath through the interactions with RH neutrinos or  the Higgs boson. 

We have demonstrated that the   primordial abundance  has sizable modification to $\Delta N_{\rm eff}$ that can be probed by the current and future CMB experiments. 
If the majoron decays relativistically, the contribution to $\Delta N_{\rm eff}$ was shown in Fig~\ref{fig:Neff_Relativistic}, which only depends on the freeze-out temperature. Things become more interesting if the majoron decays in the nonrelativistic regime. As was shown in Fig.~\ref{fig:Neff_CMB}, the current Planck measurements could already put severe constraints on the lepton-number breaking scale. More importantly, in this case, a larger $f$ would lead to a larger $\Delta N_{\rm eff}$. Therefore, opposite to the freeze-in scenario, upper bounds of $f$ are obtained from   primordial majoron decay. Furthermore, when  both the freeze-out and freeze-in abundances coexist, we obtain a sandwiched window for $f$ in terms of the majoron mass. 

Given the forecast sensitivities of future SO and CMB-S4 experiments, we showed the sandwiched windows in Fig.~\ref{fig:Neff_SO_CMBS4}. It can be seen that null signals of $\Delta N_{\rm eff}$ in future CMB experiments will push the low-scale seesaw scenario into a narrow sandwiched parameter space. In particular, if the primordial majoron abundance is inherited from relativistic freeze-out and later decays nonrelativistically into neutrinos, null signals of $\Delta N_{\rm eff}$ from CMB-S4 is able to fully close such a low-scale seesaw window.
 
Finally, although we have mainly focused on the majoron abundance in this work, it is worthwhile to emphasize that any new light particle coupled to neutrinos or photons can also be expected to result in a sandwiched window in the parameter space, as long as the particle has abundances from both the UV sources and the freeze-in production and only decays after the neutrino decoupling. Such a general phenomenon deserves careful investigation in the future.

\begin{acknowledgments}
We would like to thank Xun-Jie Xu for helpful discussions. S.-P.~Li  is supported in part by the National Natural Science Foundation of China under grant No.~12141501 and also by CAS Project for Young Scientists in Basic Research (YSBR-099). BY is supported in part by the Natural Science Foundation of China under grant No.~11835013.
\end{acknowledgments}

\appendix

\section{Majoron interactions}
\label{appen:majoron-interection}
In this appendix, we summarize the majoron interactions in the singlet majoron model. Most of the results in this section are not new and have been given in the literature but without details~\cite{Schechter:1981cv,Pilaftsis:1993af}. Nevertheless, we would like to derive them here for the purpose of completeness and self-consistency, and also to set up our notations.

\subsection{Majoron-neutrino interactions}\label{appen:majoron-neutrino}
First of all, 
before the electroweak gauge symmetry breaking,  the majoron only couples to RH neutrinos. In this case, one can take the basis where the mass matrix of RH neutrinos is diagonal, i.e., $Y_N=\widehat{Y}_N\equiv {\rm Diag}\left(Y_1,Y_2,Y_3\right)$. Defining the mass eigenstates $\widehat{N}\equiv N_{\rm R}+N_{\rm R}^c=\left(N_1,N_2,N_3\right)^{\rm T}$, we can write $N_{\rm R}=P_{\rm R} \widehat{N}$ and $N_{\rm R}^c = P_{\rm L}\widehat{N}$ with $P_{\rm R/L}\equiv \left(1\pm\gamma_5\right)/2$ being the projection operators. Then we have
\begin{eqnarray}
	\overline{N_{\rm R}^c} N_{\rm R} = \overline{\widehat{N}}P_{\rm R}\widehat{N}\,,\qquad
	\overline{N_{\rm R}}N_{\rm R}^c = \overline{\widehat{N}}P_{\rm L}\widehat{N}\,.
\end{eqnarray}
Therefore, from Eq.~(\ref{eq:Lag}), we arrive at the interaction between the majoron and RH neutrinos:
\begin{eqnarray}
	\label{eq:JNNdiag}
	{\cal L}_J = -\frac{{\rm i} J}{2\sqrt{2}}\overline{N_{\rm R}^c}\widehat{Y}_N N_{\rm R} + \frac{{\rm i} J}{2\sqrt{2}}\overline{N_{\rm R}}\widehat{Y}_N N_{\rm R}^c
	= -\frac{{\rm i} J}{2\sqrt{2}}\sum_{i=1}^{3}Y_i \overline{N_i} \gamma_5 N_i\, .
\end{eqnarray}

After the electroweak gauge symmetry breaking, the Higgs boson acquires the non-zero VEV and the majoron will interact with active neutrinos through the flavor mixing between active and sterile neutrinos. The relevant Lagrangian reads
\begin{eqnarray}
	\label{eq:Yukawa2}
	{\cal L} &=& -\overline{\nu_{\rm L}} M_{\rm D} N_{\rm R}-\frac{1}{2} \overline{N_{\rm R}^c}M_{\rm R} N_{\rm R}-\frac{{\rm i} J}{2\sqrt{2}}\overline{N_{\rm R}^c}Y_N N_{\rm R}+{\rm h.c.}
	\nonumber\\
	&=& -\frac{1}{2}
	\begin{pmatrix}
		\overline{\nu_{\rm L}}& \overline{N_{\rm R}^c}
	\end{pmatrix}
	\begin{pmatrix}
		0&M_{\rm D}\\
		M_{\rm D}^{\rm T}&M_{\rm R}
	\end{pmatrix}
	\begin{pmatrix}
		\nu_{\rm L}^c\\
		N_{\rm R}
	\end{pmatrix}
	-\frac{{\rm i} J}{2\sqrt{2}}\overline{N_{\rm R}^c}Y_N N_{\rm R}+{\rm h.c.}\,,
\end{eqnarray}
where $M_{\rm D}=Y_\nu v/\sqrt{2}$ is the Dirac mass matrix. The $6\times 6$ mass matrix in Eq.~(\ref{eq:Yukawa2}) can be diagonalized via
\begin{eqnarray}
	\label{eq:diagonalization}
	{\cal U}^\dagger \begin{pmatrix}
		0&M_{\rm D}\\
		M_{\rm D}^{\rm T}&M_{\rm R}
	\end{pmatrix} {\cal U}^{*} =
	\begin{pmatrix}
		\widehat{m}&0\\
		0&\widehat{M}
	\end{pmatrix},\qquad
	{\cal U}\equiv
	\begin{pmatrix}
		V&R\\
		S&U
	\end{pmatrix},
\end{eqnarray} 
where $\widehat{m}={\rm Diag}\left(m_1,m_2,m_3\right)$ and $\widehat{M}={\rm Diag}\left(M_1,M_2,M_3\right)$ are the diagonal mass matrices for light and heavy neutrinos, respectively. Here we have explicitly split the $6\times 6$ unitary matrix ${\cal U}$ into four $3\times 3$ sub-matrices $V$, $R$, $S$ and $U$. From Eq.~(\ref{eq:diagonalization}) and the unitarity of ${\cal U}$, one can obtain the following conditions
\begin{eqnarray}
	&&V^\dagger V + S^\dagger S = U^\dagger U + R^\dagger R = {\bf 1}\,,\label{eq:condition1}\\
	&&V^\dagger R + S^\dagger U = {\bf 0}\,,\label{eq:condition2}\\
	&&V \widehat{m} V^{\rm T} + R \widehat{M} R^{\rm T} = {\bf 0}\,,\label{eq:condition3}
\end{eqnarray}
which will be useful later. Changing from the flavor eigenstates to the mass eigenstates via
\begin{eqnarray}
	\nu_{\rm L} &=& V \widehat{\nu}_{\rm L} + R \widehat{N}_{\rm R}^c\,,\nonumber\\
	N_{\rm R} &=& S^{*} \widehat{\nu}_{\rm L}^c + U^{*} \widehat{N}_{\rm R}\,,
\end{eqnarray}
the mass term in Eq.~(\ref{eq:Yukawa2}) becomes
\begin{eqnarray}
	{\cal L}_{\rm mass} &=& -\frac{1}{2}
	\begin{pmatrix}
		\overline{\widehat{\nu}_{\rm L}}& \overline{\widehat{N}_{\rm R}^c}
	\end{pmatrix}
	\begin{pmatrix}
		\widehat{m}&0\\
		0&\widehat{M}
	\end{pmatrix}
	\begin{pmatrix}
		\widehat{\nu}_{\rm L}^c\\
		\widehat{N}_{\rm R}
	\end{pmatrix}+{\rm h.c.}
\nonumber\\
	&=&-\frac{1}{2}\sum_{i=1}^{3}\left(m_i \overline{\nu_i}\nu_i + M_i \overline{N_i} N_i\right),
\end{eqnarray} 
where we have defined $\widehat{\nu}_{\rm L} + \widehat{\nu}_{\rm L}^c \equiv \widehat{\nu} = \left(\nu_1, \nu_2, \nu_3\right)^{\rm T}$ and $\widehat{N}_{\rm R} + \widehat{N}_{\rm R}^c \equiv \widehat{N} = \left(N_1, N_2, N_3\right)^{\rm T}$. In the meanwhile, the interaction term in Eq.~(\ref{eq:Yukawa2}) becomes
\begin{align}
	\label{eq:LJ}
	{\cal L}_J &= -\frac{{\rm i}J}{2f}\left(\overline{\widehat{\nu}_{\rm L}} S^\dagger +\overline{\widehat{N}_{\rm R}^c} U^\dagger\right)\left(S \widehat{m} S^{\rm T} + U \widehat{M} U^{\rm T}\right)\left(S^{*} \widehat{\nu}_{\rm L}^c + U^{*} \widehat{N}_{\rm R}\right) + {\rm h.c.}
	\nonumber\\
	&=-\frac{{\rm i}J}{2f}
	\begin{pmatrix}
		\overline{\widehat{\nu}_{\rm L}}& \overline{\widehat{N}_{\rm R}^c}
	\end{pmatrix}
	\begin{pmatrix}
		S^\dagger S & S^\dagger U\\
		U^\dagger S & U^\dagger U
	\end{pmatrix}
	\begin{pmatrix}
		\widehat{m}&0\\
		0&\widehat{M}
	\end{pmatrix}
	\begin{pmatrix}
		S^\dagger S & S^\dagger U\\
		U^\dagger S & U^\dagger U
	\end{pmatrix}^{*}
	\begin{pmatrix}
		\widehat{\nu}_{\rm L}^c\\
		\widehat{N}_{\rm R}
	\end{pmatrix}+{\rm h.c.}
\nonumber\\
	&=-\frac{{\rm i}J}{2f}
	\begin{pmatrix}
		\overline{\widehat{\nu}_{\rm L}}& \overline{\widehat{N}_{\rm R}^c}
	\end{pmatrix}\left[
	\begin{pmatrix}
		{\bf 1} & {\bf 0}\\
		{\bf 0} & {\bf 1}
	\end{pmatrix}-
	\begin{pmatrix}
		V^\dagger V & V^\dagger R\\
		R^\dagger V & R^\dagger R
	\end{pmatrix}
	\right]
	\begin{pmatrix}
		\widehat{m}&0\\
		0&\widehat{M}
	\end{pmatrix}\nonumber\\
&\qquad\qquad\qquad\;\; \times
	\left[
	\begin{pmatrix}
		{\bf 1} & {\bf 0}\\
		{\bf 0} & {\bf 1}
	\end{pmatrix}-
	\begin{pmatrix}
		V^\dagger V & V^\dagger R\\
		R^\dagger V & R^\dagger R
	\end{pmatrix}^{*}
	\right]
	\begin{pmatrix}
		\widehat{\nu}_{\rm L}^c\\
		\widehat{N}_{\rm R}
	\end{pmatrix}+{\rm h.c.}\,,
\end{align}
where in the third line we have used the unitarity conditions in Eqs.~(\ref{eq:condition1}) and (\ref{eq:condition2}). To simplify the notation, it is helpful to define $n_{\rm L} \equiv \left(\widehat{\nu}_{\rm L}, \widehat{N}_{\rm R}^c\right)^{\rm T}$ and $n_{\rm L} + n_{\rm L}^c \equiv n = \left(n_1,n_2,...,n_6\right)^{\rm T}$, where $n_i$ (for $i=1,2,3$) correspond to the mass eigenstates of active neutrinos while $n_i$ (for $i=4,5,6$) correspond to those of sterile neutrinos. Then it follows that $\overline{n_{\rm L}}n_{\rm L}^c=\overline{n}P_{\rm R}n$ and $\overline{n_{\rm L}^c}n_{\rm L}=\overline{n}P_{\rm L}n$.
In addition, we define
\begin{eqnarray}
	{\cal C} \equiv 
	\begin{pmatrix}
		V^\dagger V & V^\dagger R\\
		R^\dagger V & R^\dagger R
	\end{pmatrix}, \qquad
	\widehat{\cal M} \equiv 
	\begin{pmatrix}
		\widehat{m}&0\\
		0&\widehat{M}
	\end{pmatrix} \equiv {\rm Diag}\left(m_1,m_2,...,m_6\right).
\end{eqnarray}
Note that the $(i,j)$ element of ${\cal C}$ can be related to the elements of the unitary matrix ${\cal U}$ via
\begin{eqnarray}
	\label{eq:Cij}
	{\cal C}_{ij} = \sum_{k=1}^{3}{\cal U}_{ki}^{*}{\cal U}_{kj}\,,\qquad
	i,j=1,2,...,6\,.
\end{eqnarray}
With above notations, Eq.~(\ref{eq:LJ}) can be reduced to
\begin{eqnarray}
	{\cal L}_J = -\frac{{\rm i}J}{2f}\overline{n_{\rm L}}\left[\widehat{{\cal M}}-\left({\cal C}\widehat{{\cal M}}+\widehat{{\cal M}}{\cal C}^{*}\right)+{\cal C}\widehat{{\cal M}}{\cal C}^{*}\right]n_{\rm L}^{c}+{\rm h.c.}
\end{eqnarray}
Using Eq.~(\ref{eq:condition3}), it is straightforward to obtain ${\cal C}\widehat{{\cal M}}{\cal C}^{*}={\bf 0}$. Therefore, we are left with
\begin{eqnarray}
	\label{eq:LJ2}
	{\cal L}_J &=& -\frac{{\rm i}J}{2f}\left[\left(\overline{n_{\rm L}}\widehat{{\cal M}}n_{\rm L}^c-\overline{n_{\rm L}^c}\widehat{{\cal M}}n_{\rm L}\right)-\left(\overline{n_{\rm L}}\left({\cal C}\widehat{{\cal M}}+\widehat{{\cal M}}{\cal C}^{*}\right)n_{\rm L}^c-\overline{n_{\rm L}^c}\left({\cal C}^{*}\widehat{{\cal M}}+{\widehat{\cal M}}{\cal C}\right)n_{\rm L}\right)\right]\nonumber\\
	&=&-\frac{{\rm i}J}{2f}\sum_{i,j=1}^{6}\left[m_i\overline{n_i}\gamma_5 n_i-\left({\cal C}\widehat{{\cal M}}+\widehat{{\cal M}}{\cal C}^{*}\right)_{ij}\overline{n_i}P_{\rm R}n_j + \left({\cal C}^{*}\widehat{{\cal M}}+\widehat{{\cal M}}{\cal C}\right)_{ij}\overline{n_i}P_{\rm L}n_j\right]\nonumber\\
	&=& -\frac{{\rm i}J}{2f}\sum_{i,j=1}^{6}\left[m_i\overline{n_i}\gamma_5 n_i - \left(m_i + m_j\right){\rm Re}\,{\cal C}_{ij}\, \overline{n_i}\gamma_5 n_j+{\rm i}\left(m_i-m_j\right){\rm Im}\,{\cal C}_{ij}\,\overline{n_i}n_j\right]\nonumber\\
	&=& -\frac{{\rm i}J}{2f}\sum_{i,j=1}^{6} \overline{n_i}\left[\gamma_5\left(m_i+m_j\right)\left(\frac{1}{2}\delta_{ij}-{\rm Re}\,{\cal C}_{ij}\right)+{\rm i}\left(m_i-m_j\right){\rm Im}\,{\cal C}_{ij}\right]n_j\,,
\end{eqnarray}
which is the general majoron-neutrino interaction given in Eq.~(\ref{eq:LJnuij}). Below we consider some special scenarios of the general interaction:
\begin{itemize}
	\item For $i,j=1,2,3$, from Eq.~(\ref{eq:Cij}) we have ${\cal C}_{ij} = \delta_{ij} - {\cal O}\left(M_{\rm D}^2/M_{\rm R}^2\right)\simeq \delta_{ij}$, leading to
	\begin{eqnarray}
		{\cal L}_{J\nu\nu}\simeq \frac{{\rm i} J}{2f}\sum_{i=1}^3 m_i\overline{\nu_i}\gamma_5 \nu_i\,.
	\end{eqnarray}
	Therefore, the interaction between the majoron and the active neutrinos is approximately diagonal and is suppressed by $m_i/f$.
	
	\item For $i,j=4,5,6$, we have ${\cal C}_{ij}\sim {\cal O}\left(M_{\rm D}^2/M_{\rm R}^2\right)$, therefore
	\begin{eqnarray}
		{\cal L}_{JNN}\simeq -\frac{{\rm i}J}{2f}\sum_{i=1}^{3}M_i \overline{N_i}\gamma_5 N_i\,,
	\end{eqnarray}
	which means the interaction between the majoron and the sterile neutrinos is also approximately diagonal.
	
	\item For $i=1,2,3$ and $j=4,5,6$, we have ${\cal C}_{ij} \sim {\cal O}\left(M_{\rm D}/M_{\rm R}\right)$ and
	\begin{eqnarray}\label{eq:JNnu}
		{\cal L}_{J\nu N}\simeq \frac{{\rm i}J}{2f}\sum_{i=1}^{3}\sum_{j=4}^{6} m_j \overline{n_i}\left(\gamma_5\,{\rm Re}\,{\cal C}_{ij}+{\rm i}\,{\rm Im}\,{\cal C}_{ij}\right)n_j\,.
	\end{eqnarray}
	This describes the interaction among the majoron, the active neutrinos and the sterile neutrinos.
\end{itemize}
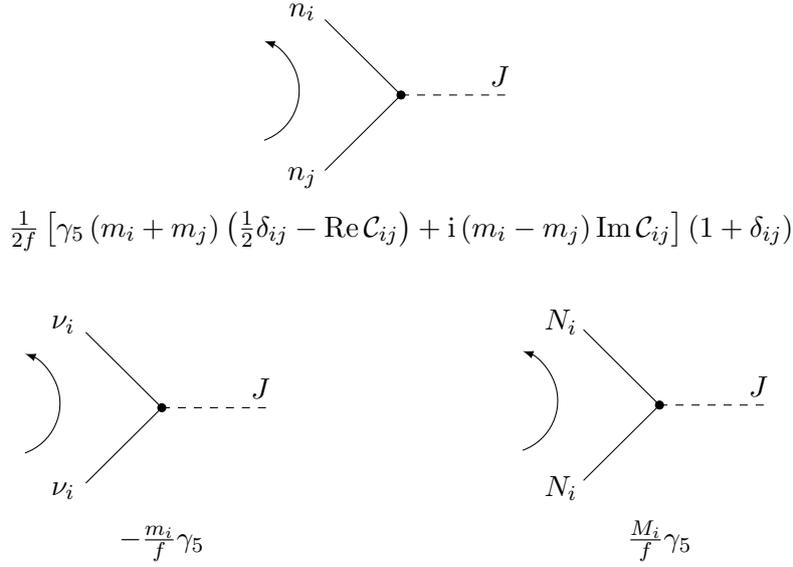
\begin{figure}[t!]
	\centering
	\begin{tikzpicture}
		\draw[black] (-1,1) -- (0,0);
		\draw[black] (-1,-1) -- (0,0);
		\draw[black,dashed] (0,0) -- (1.4,0);
		\filldraw[black] (0,0) circle (1.5pt);
		\node at (0,-1.8) {$\frac{1}{2f}\left[\gamma_5\left(m_i+m_j\right)\left(\frac{1}{2}\delta_{ij}-{\rm Re}\,{\cal C}_{ij}\right)+{\rm i}\left(m_i-m_j\right){\rm Im}\,{\cal C}_{ij}\right]\left(1+\delta_{ij}\right)$};
		\node (a) at (-1.3,+1.1) {$n_i$};
		\node (b) at (-1.3,-1.1) {$n_j$};
		\node at (1.3,0.25) {$J$};
		\draw[-latex,black] ($(b)+(-0.5,0.5)$) arc
		[
		start angle=-70,
		end angle=70,
		x radius=0.7cm,
		y radius =0.7cm
		];
	\end{tikzpicture}\\
\vspace{0.5cm}
	\begin{tikzpicture}
	\draw[black] (-1,1) -- (0,0);
	\draw[black] (-1,-1) -- (0,0);
	\draw[black,dashed] (0,0) -- (1.4,0);
	\filldraw[black] (0,0) circle (1.5pt);
	\node at (0,-1.8) {$-\frac{m_i}{f}\gamma_5$};
	\node (a) at (-1.3,+1.1) {$\nu_i$};
	\node (b) at (-1.3,-1.1) {$\nu_i$};
	\node at (1.3,0.25) {$J$};
	\draw[-latex,black] ($(b)+(-0.5,0.5)$) arc
	[
	start angle=-70,
	end angle=70,
	x radius=0.7cm,
	y radius =0.7cm
	] ;
\end{tikzpicture}
\qquad\qquad\qquad\qquad
	\begin{tikzpicture}
	\draw[black] (-1,1) -- (0,0);
	\draw[black] (-1,-1) -- (0,0);
	\draw[black,dashed] (0,0) -- (1.4,0);
	\filldraw[black] (0,0) circle (1.5pt);
	\node at (0,-1.8) {$\frac{M_i}{f}\gamma_5$};
	\node (a) at (-1.3,+1.1) {$N_i$};
	\node (b) at (-1.3,-1.1) {$N_i$};
	\node at (1.3,0.25) {$J$};
	\draw[-latex,black] ($(b)+(-0.5,0.5)$) arc
	[
	start angle=-70,
	end angle=70,
	x radius=0.7cm,
	y radius =0.7cm
	] ;
\end{tikzpicture}
\caption{\label{fig:majoron-neutrino}Feynman rules for the majoron-neutrino interactions. The arc arrows denote the orientations of fermion flow~\cite{Denner:1992vza}. 
We have identified $\nu_i\equiv n_i$ and $N_{i}\equiv n_{i+3}$ (for $i=1,2,3$). Note that when $i=j$, there is a factor of $2!$ due to the  two  indistinguishable neutrinos. This is why there is an additional factor of $(1+\delta_{ij})$ compared with Eq.~(\ref{eq:LJ2}).}
\end{figure}
The corresponding Feynman rules for majoron-neutrino interactions are summarized in Fig.~\ref{fig:majoron-neutrino}.

\subsection{Majoron-Higgs interactions}\label{appen:majoron-Higgs}
When extended with a complex singlet $S$, the general scalar potential which obeys the gauge symmetry and global ${\rm U}(1)_{L}$  symmetry is given by
\begin{align}
	\label{eq:potential}
	V\left(\Phi,S\right)=\mu_\Phi^2 \Phi^\dagger \Phi + \mu_S^2S^\dagger S+\frac{\lambda_\Phi}{2}\left(\Phi^\dagger \Phi\right)^2 + \frac{\lambda_S}{2}\left(S^\dagger S\right)^2+\lambda_{\Phi S} \left(\Phi^\dagger \Phi\right)\left(S^\dagger S\right),
\end{align}
where
\begin{eqnarray}
	\label{eq:field}
	\Phi = \begin{pmatrix}
		G^{+}\\
		\frac{v}{\sqrt{2}}+\frac{h'+{\rm i}G^0}{\sqrt{2}}
	\end{pmatrix},\qquad
	S=\frac{f}{\sqrt{2}}+\frac{\rho'+{\rm i}J}{\sqrt{2}}\, ,
\end{eqnarray}
with $\Phi$ the SM Higgs doublet.
Substituting the VEVs into the potential we have
\begin{eqnarray}
	V\left(v,f\right)\equiv V\left(\langle \Phi \rangle,\langle S \rangle\right)=\frac{1}{2} \mu_\Phi^2 v^2 + \frac{1}{2} \mu_S^2 f^2 + \frac{\lambda_\Phi^2}{8} v^4 + \frac{\lambda_S^2}{8} f^4 + \frac{\lambda_{\Phi S}}{4} v^2 f^2\,.
\end{eqnarray}
The VEVs are determined by minimizing the scalar potential,
\begin{eqnarray}
	\frac{1}{v}\frac{\partial}{\partial v} V\left(v,f\right) &=& \mu_\Phi^2 +\frac{\lambda_\Phi}{2}v^2+\frac{\lambda_{\Phi S}}{2} f^2=0\,,\\
	\frac{1}{f}\frac{\partial}{\partial f} V\left(v,f\right) &=& \mu_S^2 +\frac{\lambda_S}{2}f^2+\frac{\lambda_{\Phi S}}{2}v^2 =0\,,
\end{eqnarray}
from which we obtain
\begin{eqnarray}
	v &=& \sqrt{\frac{2\left(\lambda_{\Phi S} \mu_S^2-\lambda_S \mu_\Phi^2\right)}{\lambda_\Phi^{} \lambda_S^{}-\lambda_{\Phi S}^2}}\,,\label{eq:VEV1}\\
	f &=& \sqrt{\frac{2\left(\lambda_{\Phi S} \mu_\Phi^2-\lambda_\Phi \mu_S^2\right)}{\lambda_\Phi^{} \lambda_S^{}-\lambda_{\Phi S}^2}}\,.\label{eq:VEV2}
\end{eqnarray}
Note that in the limit of $\lambda_{\Phi S} \to 0$, the VEVs reduce to $v=\sqrt{-2\mu_\Phi^2/\lambda_\Phi}$ and $f=\sqrt{-2\mu_S^2/\lambda_S}$.

To calculate the spectrum of scalars in the singlet majoron model, we substitute Eq.~(\ref{eq:field}) into Eq.~(\ref{eq:potential}) and keep only quadratic terms, obtaining
\begin{eqnarray}
	\label{eq:mass}
	V \supset \frac{\lambda_\Phi}{2}v^2 h'^2 + \frac{\lambda_S}{2}f^2 \rho'^2 + \lambda_{\Phi S} v f h' \rho' = \frac{1}{2}
	\begin{pmatrix}
		h'&\rho'
	\end{pmatrix}
	\begin{pmatrix}
		\lambda_\Phi v^2 & \lambda_{\Phi S} v f\\
		\lambda_{\Phi S} v f & \lambda_S f^2
	\end{pmatrix}
	\begin{pmatrix}
		h'\\
		\rho'
	\end{pmatrix},
\end{eqnarray}
where Eqs.~(\ref{eq:VEV1}) and (\ref{eq:VEV2}) have been used to eliminate the quadratic terms of $G^{\pm}$, $G^0$ and $J$. Therefore, we are left with four Nambu-Goldstone bosons
\begin{eqnarray}
	m_{G^{\pm}} = m_{G^{0}} = m_J =0\,.
\end{eqnarray}
Among them,   $G^{\pm}$ and $G^{0}$ will be absorbed into the longitudinal components of gauge bosons, while $J$ will be left massless because it does not participate in any gauge interaction. 

The mass of $J$ can be generated, for example, by introducing terms that explicitly break the  global ${\rm U}(1)_{L}$  symmetry~\cite{Langacker:1986rj,Akhmedov:1992hi,Rothstein:1992rh,Gu:2010ys,Frigerio:2011in}.  A simple way for generating the majoron mass is to   add a tree-level  ${\rm U}(1)_L$-breaking term:
\begin{align}
\lambda_{b}S^2\Phi^\dagger \Phi\,,
\end{align} 
such that the majoron gets a tree-level mass $m_J^2=\lambda_{b}v^2/2$. For $m_J<1$~MeV, we arrive at $\lambda_{b}<10^{-11}$. It should be mentioned that such a small coupling cannot help to thermalize the majoron, as was pointed below Eq.~\eqref{eq:Higgs_portal}. In addition, while the majoron abundance can be generated by the freeze-in Higgs decay $h\to 2J$, it is much suppressed with respect to the neutrino coalescence $2\nu\to J$,  as the former already eases at $T\simeq \mathcal{O}(m_h)\gg m_J$. Therefore, the impacts on the cosmological evolution of the majoron from this simple case can be safely neglected.

In addition, the mass eigenstates of the two CP-even neutral scalars $\left\{h,\rho\right\}$ are given by diagonalizing the mass matrix in Eq.~(\ref{eq:mass}),
\begin{eqnarray}
	\begin{pmatrix}
		h\\
		\rho
	\end{pmatrix}=
	\begin{pmatrix}
		\cos\theta & \sin\theta\\
		-\sin\theta & \cos\theta
	\end{pmatrix}
	\begin{pmatrix}
		h'\\
		\rho'
	\end{pmatrix},
\end{eqnarray}
where the mixing angle results from the Higgs portal coupling $\lambda_{\Phi S}$, with 
\begin{eqnarray}
	\tan 2\theta = \frac{2\lambda_{\Phi S} \tan\beta}{\lambda_\Phi ^2 \tan^2 \beta - \lambda_S^2}\qquad (\tan\beta\equiv \frac{v}{f})\, .
\end{eqnarray}
Moreover, the masses turn out to be
\begin{eqnarray}
	m_h &=& \sqrt{\lambda_\Phi v^2 \cos^2\theta+\lambda_S f^2 \sin^2\theta + \lambda_{\Phi S} v f \sin2\theta}\,,\\
	m_\rho &=& \sqrt{\lambda_S f^2 \cos^2\theta+\lambda_\Phi v^2 \sin^2\theta - \lambda_{\Phi S} v f \sin2\theta}\,.
\end{eqnarray}
In the limit of $\lambda_{\Phi S}\to 0$ we have $m_h=\sqrt{\lambda_\Phi} v$ and $m_\rho = \sqrt{\lambda_{S}} f$.

\begin{figure}[t!]
	\centering
	\begin{tikzpicture}
		\draw[black, dashed] (-1,1) -- (1,-1);
		\draw[black, dashed] (-1,-1) -- (1,1);
		\filldraw[black] (0,0) circle (1.5pt);
		\node at (0,-1.8) {$-{\rm i}\,3\lambda_S$};
		\node at (-1.3,-1.1) {$J$};
		\node at (-1.3,+1.1) {$J$};
		\node at (+1.3,-1.1) {$J$};
		\node at (+1.3,+1.1) {$J$};
	\end{tikzpicture}\qquad
	\begin{tikzpicture}
		\draw[black, dashed] (-1,1) -- (1,-1);
		\draw[black, dashed] (-1,-1) -- (1,1);
		\filldraw[black] (0,0) circle (1.5pt);
		\node at (0,-1.8) {$-{\rm i}\,\lambda_{\Phi S}$};
		\node at (-1.3,-1.1) {$G^0$};
		\node at (-1.3,+1.1) {$J$};
		\node at (+1.3,-1.1) {$G^0$};
		\node at (+1.3,+1.1) {$J$};
	\end{tikzpicture}\qquad
	\begin{tikzpicture}
		\begin{scope}[decoration={markings,mark=at position 0.6 with {\arrow{latex}}}] 
			\draw[black, dashed,postaction={decorate}] (0,0) -- (1,-1);
		\end{scope}
		\begin{scope}[decoration={markings,mark=at position 0.5 with {\arrow{latex}}}] 
			\draw[black, dashed,postaction={decorate}] (-1,-1) -- (0,0);
		\end{scope}
		\draw[black, dashed] (-1,1) -- (0,0);
		\draw[black, dashed] (0,0) -- (1,1);
		\filldraw[black] (0,0) circle (1.5pt);
		\node at (0,-1.8) {$-{\rm i}\,\lambda_{\Phi S}$};
		\node at (-1.3,-1.1) {$G^-$};
		\node at (-1.3,+1.1) {$J$};
		\node at (+1.3,-1.1) {$G^+$};
		\node at (+1.3,+1.1) {$J$};
	\end{tikzpicture}\\
	\vspace{1cm}
	\begin{tikzpicture}
		\draw[black, dashed] (-1,1) -- (1,-1);
		\draw[black, dashed] (-1,-1) -- (1,1);
		\filldraw[black] (0,0) circle (1.5pt);
		\node at (0,-1.8) {$-{\rm i}\left(\lambda_S \cos^2\theta+\lambda_{\Phi S}\sin^2\theta\right)$};
		\node at (-1.3,-1.1) {$\rho$};
		\node at (-1.3,+1.1) {$J$};
		\node at (+1.3,-1.1) {$\rho$};
		\node at (+1.3,+1.1) {$J$};
	\end{tikzpicture}\qquad
	\begin{tikzpicture}
		\draw[black, dashed] (-1,1) -- (0,0);
		\draw[black, dashed] (1,1) -- (0,0);
		\draw[black,dashed] (0,0) -- (0,-1.4);
		\filldraw[black] (0,0) circle (1.5pt);
		\node at (0,-1.8) {$-{\rm i}\left(\lambda_S f\cos\theta-\lambda_{\Phi S}v\sin\theta\right)$};
		\node at (-1.3,+1.1) {$J$};
		\node at (+1.3,+1.1) {$J$};
		\node at (0.2,-1.3) {$\rho$};
	\end{tikzpicture}\\
	\vspace{1cm}
	\begin{tikzpicture}
		\draw[black, dashed] (-1,1) -- (1,-1);
		\draw[black, dashed] (-1,-1) -- (1,1);
		\filldraw[black] (0,0) circle (1.5pt);
		\node at (0,-1.8) {$-{\rm i}\left(\lambda_{\Phi S} \cos^2\theta+\lambda_{S}\sin^2\theta\right)$};
		\node at (-1.3,-1.1) {$h$};
		\node at (-1.3,+1.1) {$J$};
		\node at (+1.3,-1.1) {$h$};
		\node at (+1.3,+1.1) {$J$};
	\end{tikzpicture}\qquad
	\begin{tikzpicture}
		\draw[black, dashed] (-1,1) -- (0,0);
		\draw[black, dashed] (1,1) -- (0,0);
		\draw[black,dashed] (0,0) -- (0,-1.4);
		\filldraw[black] (0,0) circle (1.5pt);
		\node at (0,-1.8) {$-{\rm i}\left(\lambda_{\Phi S} v\cos\theta+\lambda_{S}f\sin\theta\right)$};
		\node at (-1.3,+1.1) {$J$};
		\node at (+1.3,+1.1) {$J$};
		\node at (0.2,-1.3) {$h$};
	\end{tikzpicture}\\
	\vspace{1cm}
	\begin{tikzpicture}
		\draw[black, dashed] (-1,1) -- (1,-1);
		\draw[black, dashed] (-1,-1) -- (1,1);
		\filldraw[black] (0,0) circle (1.5pt);
		\node at (0,-1.8) {$-{\rm i}\,\frac{1}{2}\left(\lambda_S-\lambda_{\Phi S}\right)\sin2\theta$};
		\node at (-1.3,-1.1) {$h$};
		\node at (-1.3,+1.1) {$J$};
		\node at (+1.3,-1.1) {$\rho$};
		\node at (+1.3,+1.1) {$J$};
	\end{tikzpicture}\\
	\vspace{0.5cm}
	\caption{\label{fig:majoron-Higgs}Feynman rules for the majoron self interaction and majoron-Higgs interactions, where $\theta$ is the mixing angle between two CP-even scalars $\rho$ and $h$.}
\end{figure}
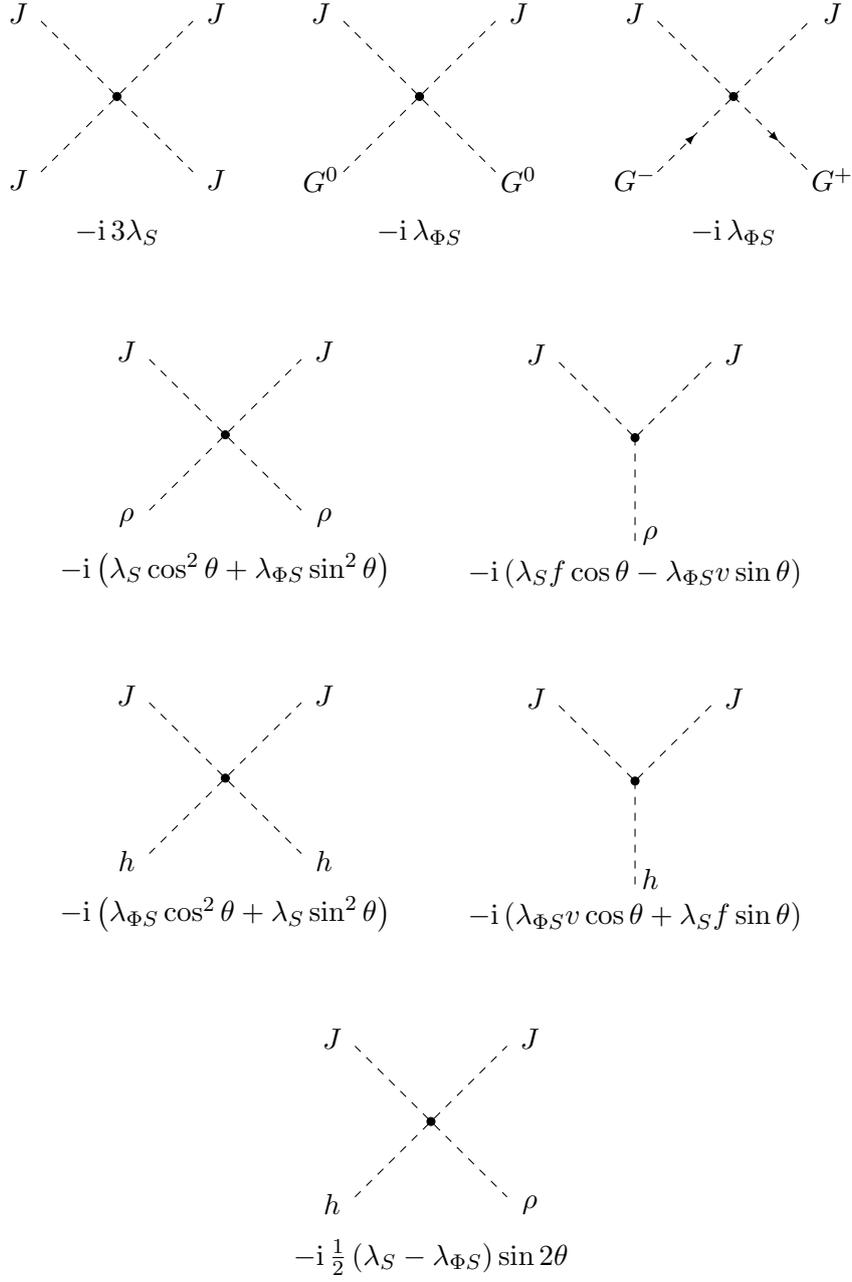

The majoron self-interaction comes from the term proportional to $\lambda_S$, while the interaction between the majoron and the Higgs field comes from the term proportional to $\lambda_{\Phi S}$. More explicitly, we have
\begin{align}
	\label{eq:Jscalarint}
	{\cal L}_J =& -\frac{\lambda_S}{8} \left(J^4+2J^2 \rho'^2+4fJ^2\rho'\right)-\frac{\lambda_{\Phi  S}}{4}\left(2J^2 G^{-}G^{+}+J^2 \left(G^0\right)^2 + J^2 h'^2 + 2v J^2 h'\right)\nonumber\\
	= &-\frac{\lambda_S}{8}J^4 - \frac{\lambda_{\Phi S}}{2}J^2 G^{-} G^{+} - \frac{\lambda_{\Phi S}}{4} J^2 \left(G^0\right)^2\nonumber\\
	& -\frac{1}{4}\left(\lambda_S \cos^2\theta + \lambda_{\Phi S}\sin^2\theta\right)J^2 \rho^2 -\frac{1}{2}\left(\lambda_S f \cos\theta-\lambda_{\Phi S} v \sin\theta\right)J^2 \rho\nonumber\\
	&  -\frac{1}{4}\left(\lambda_{\Phi S} \cos^2\theta + \lambda_S\sin^2\theta\right)J^2 h^2 -\frac{1}{2}\left(\lambda_{\Phi S} v \cos\theta+\lambda_S f \sin\theta\right)J^2 h\nonumber\\
	& -\frac{1}{4}\left(\lambda_S-\lambda_{\Phi S}\right)\sin2\theta J^2 \rho h\;.
\end{align} 
The Feynman rules for the majoron self-interaction and majoron-Higgs interactions are summarized in Fig.~\ref{fig:majoron-Higgs}.

\section{Neutrino coalescence rate}\label{appen:Cn_2vJ}
The collision term of the neutrino coalescence $2\nu \to J$ in Eq.~(\ref{eq:dYJdT-fi}) is given by
\begin{align}
	\mathcal{C}_{2\nu\to J}^{n} & =\int\frac{{\rm d}^{3}p}{(2\pi)^{3}2E}\frac{{\rm d}^{3}p_{1}}{(2\pi)^{3}2E_{1}}\frac{{\rm d}^{3}p_{2}}{(2\pi)^{2}2E_{2}}(2\pi)^{4}\delta^{4}|\mathcal{M}|_{2\nu\to J}^{2}f_{1}(E_{1})f_{2}(E_{2})\,,\label{eq:Cn_nunuJ}
\end{align}
where $\delta^4\equiv \delta^{4}(p_{1}+p_{2}-p)$, and the quantum statistics for $J$ is neglected. 
$f_{i}(E_{i})=(e^{E_{i}/T_{\nu}}+1)^{-1}$ are distribution functions of incoming neutrinos, 
with $T_{\nu}$ the neutrino temperature, and the squared
amplitude is given by 
\begin{align}
	|\mathcal{M}|_{2\nu\to J}^{2} & \simeq \frac{2m_{J}^{2}}{f^{2}}\sum_{i=1}^{3}m_{i}^{2}\,.\label{eq:ampsq_nunuJ}
\end{align}
To calculate Eq.~(\ref{eq:Cn_nunuJ}), we can first go to the rest
frame of $J$. Suppose in the rest plasma frame with 4-velovity
$u^\mu=(1,\vec{0})$, the 4-momentum of the majoron is denoted by $p^{\mu}=(E,\vec{p})$.
Then changing to the rest frame of $J$ with $p'^{\mu}=(m_{J},\vec{0})$
is equivalent to boosting the rest plasma frame with $u^{\prime\mu}=(E/m_{J},-\vec{p}/m_{J}).$
In the rest frame of $J$, we denote the neutrino 4-momenta by $p'^{\mu}_{1,2}=(\omega_{1,2},\vec{q}_{1,2})$,
where $\omega_{1}=\omega_{2}=m_{J}/2, |\vec{q}_{1}|=|\vec{q}_{2}|=\beta_{\nu} m_{J}/2$ are determined by the momentum conservation, with $\beta_{\nu}\equiv\sqrt{1-4m_{i}^{2}/m_{J}^{2}}$ the neutrino velocity. Then, in the
rest frame of $J$, the collision term is given by

\begin{align}
	\mathcal{C}_{2\nu\to J}^{n} & =\int\frac{{\rm d}^{3}p}{(2\pi)^{3}2E}\frac{{\rm d}^{3}q_{1}}{(2\pi)^{3}2\omega_{1}}\frac{{\rm d}^{3}q_{2}}{(2\pi)^{2}2\omega_{2}}(2\pi)^{4}\delta^{3}(\vec{q}_{1}+\vec{q}_{2})\,\delta(2\sqrt{|\vec{q}_{1}|^{2}+m_{i}^{2}}-m_{J})  \nonumber \\[0.2cm]
	& \times |\mathcal{M}|_{2\nu\to J}^{2}f_{1}(p_{1}^{\prime}\cdot u^{\prime})f_{2}(p_{2}^{\prime}\cdot u^{\prime})\,,\label{eq:Cn_nunuJ_restframe}
\end{align}
where the distribution function of the thermalized neutrinos
in the rest frame of $J$ reads

\begin{align}
	f_{i}(p_{i}'\cdot u') & =\frac{1}{e^{p_{i}'\cdot u'/T}+1}=\frac{1}{e^{(E/2+\vec{q_{i}}\cdot\vec{p}/m_{J})/T_{\nu}}+1}\,,\label{eq:distribution_rest}
\end{align}
for $i=1,2$. Integrating out $|\vec{q}_{2}|$ and
$|\vec{q}_{1}|$ via $\delta$-functions, we are led to 

\begin{align}
	\mathcal{C}_{2\nu\to J}^{n} & =\frac{\beta_{\nu}}{64\pi^{3}}|\mathcal{M}|_{2\nu\to J}^{2}\int_0^\infty\frac{p^{2}{\rm d}p}{E}\int_{-1}^1 {\rm d}\cos\theta f_{1}(p_{1}^{\prime}\cdot u^{\prime})f_{2}(p_{2}^{\prime}\cdot u^{\prime})\nonumber \\[0.2cm]
	& =\frac{T_{\nu}}{16\pi^{3}}|\mathcal{M}|_{2\nu\to J}^{2}\int_0^\infty {\rm d}p\frac{p}{E}\frac{1}{e^{E/T_{\nu}}+1}\log\left(\frac{e^{\frac{E+\beta_{\nu}p}{2T_{\nu}}}+1}{e^{\frac{E}{2T_{\nu}}}+e^{\frac{\beta_{\nu}p}{2T_{\nu}}}}\right),\label{eq:Cn_nunuJ_fin}
\end{align}
Taking the Boltzmann approximation, we have
\begin{align}
	\mathcal{C}_{2\nu\to J}^{n} & \simeq\frac{\beta_{\nu}}{32\pi^{3}}|\mathcal{M}|_{2\nu\to J}^{2}\int_0^\infty {\rm d}E \sqrt{E^{2}-m_{J}^{2}}\,e^{-E/T_{\nu}}\simeq\frac{T_{\nu}m_{J}^{3}}{16\pi^{3}f^{2}}\sum_{i=1}^{3}m_{i}^{2}K_{1}(m_{J}/T_{\nu})\,,\label{eq:Cn_nunuJ_Boltzmann}
\end{align}
where in the last step we have assumed $m_i \ll m_J$ so that $\beta_{\nu}\simeq 1$, and $K_1$ is the modified Bessel function.


\bibliographystyle{JHEP}
\bibliography{Refs}

\end{document}